\documentclass[12pt]{article}
\usepackage{epsf}
\begin{document}

\def\be{\begin{equation}}
\def\ee{\end{equation}}
\def\ba{\begin{array}{l}}
\def\ea{\end{array}}
\def\bea{\begin{eqnarray}}
\def\eea{\end{eqnarray}}
\def\eq#1{(\ref{#1})}
\def\fig#1{Fig \ref{#1}} 
\def\del{\partial}
\def\sbh{S_{\rm BH}}
\def\bull{$\bullet$}
\def\gap{\vspace{10ex}}
\def\tgap{\vspace{3ex}}
\def\sgap{\vspace{5ex}}
\def\lgap{\vspace{20ex}}
\def\half{\frac{1}{2}}
\def\pto{\vfill\eject}
\def\gst{g_{\rm st}}
\def\tC{{\widetilde C}}
\def\z{{\bar z}}
\def\o{{\cal O}}
\def\S{{\cal S}}
\def\X{{\cal X}}
\def\N{{\cal N}}
\def\D{{\tilde D}}
\def\re#1{{\bf #1}}
\def\nn{\nonumber}
\def\nl{\hfill\break}
\def\ni{\noindent}
\def\bibi{\bibitem}
\def\c#1{{\hat{#1}}}

\pretolerance=1000000

\begin{flushright}
TIFR/TH/00-11\\
CERN-TH/2000-063\\
hep-th/0002184\\
February 2000\\
\end{flushright}
\begin{center}
\vspace{2 ex}
{\large\bf A review of the D1/D5 system and five dimensional
black hole from supergravity and brane viewpoints}
\\
\vspace{3 ex}
Gautam Mandal\\
~\\
{\sl Theory Division, CERN, CH-1211, Geneva 23, Switzerland.}\\
and \\
{\sl Department of Theoretical Physics,
Tata Institute of Fundamental Research,}\\
{\sl Homi Bhabha Road, Mumbai 400 005, INDIA.} \\
\vspace{10 ex}
\pretolerance=1000000
\bf ABSTRACT\\
\end{center}
\vspace{1 ex}

We review$^*$ some aspects of the D1/D5 system of type IIB string
theory and the associated five dimensional black hole. We include a
pedagogical discussion of the construction of relevant classical
solutions in supergravity.  We discuss the gauge theory and the
conformal field theory relevant to D-brane description of these
systems. In order to discuss Hawking radiation we are automatically
led to a discussion of near-horizon geometries and their relation to
gauge theories and conformal field theories. We show how inputs from
AdS/CFT correspondence resolve some earlier puzzles regarding Hawking
radiation. Besides the D1/D5 system, we include a brief discussion of
some nonsupersymmetric systems which show unexpected agreement between
supergravity and perturbative brane/string computations. We also
comment briefly on possible implications of the AdS/CFT relation
for the correspondence principle and for the principle of black hole
complementarity.

\vfill
\vspace{5ex}
\hrule
\begin{flushleft}
\baselineskip 2ex
{\small e-mail:  Gautam.Mandal@cern.ch, mandal@theory.tifr.res.in}\\
{\small  $*$ Expanded version of lectures presented 
at the ICTP Spring School, 
April 1999.}

\end{flushleft}
\clearpage

\vspace{8ex}

\section{\large Introduction} 

\tgap

There are a number of theoretical reasons why black holes are
important objects to study (for books discussing black holes at
various levels, see, e.g.
\cite{Mis-Tho-Whe-book,Wal-book,Haw-book,Haw-Isr-book}).  From the
point of view of classical gravity, they constitute 
a very interesting class
of spacetimes, characterized by event horizons and (often)
singularities and tied to the enigmatic  no-hair theorems
\cite{hair-Review} and the laws of black hole thermodynamics
\cite{Haw76a,Car79}.  Each of these features embodies some challenging
questions. For example,\nl
\ni(a) an event horizon implies a time-reversal
asymmetry \cite{Haw76a,Pen79}: objects can enter through the event
horizon but the time-reversed process in which they would come out, is
not allowed, at least classically;\nl
\ni (b) the presence of a spacetime
singularity typically signals a breakdown of the metric 
description and signals
some deeper underlying physics;\nl
\ni (c) the no-hair theorem appears to be in
conflict with the existence of a black hole entropy: if for a given
charge, mass and angular momentum (and possibly a few
other quantum numbers) all black holes are identical as
measured by an observer at asymptotic infinity then the number of {\em
  accessible} states possessing those quantum
numbers should be 1, implying a zero entropy;
\nl\ni (d)  the
existence of a black hole entropy proportional to the {\em area} of
the event horizon, coupled with Bekenstein's bound \cite{Bek-bound} on
maximum entropy in a given region of spacetime, appears to necessitate
a fundamental revision in our notion of degrees of freedom in the
presence of gravity.

With the introduction of quantum mechanics, black holes assume an even
more fundamental role. The temperature and entropy that appear in
classical black hole thermodynamics are shown by Hawking \cite{Haw75}
to refer to radiation coming out of a black hole, the value of the
temperature being proportional to $\hbar$. This would appear to imply
evolution of a pure state of collapsing matter to radiation, a mixed
state, at least when the black hole evaporates completely. Such an
evolution is contrary to unitarity and would signal incompatibility
between conventional quantum mechanics and general relativity
\cite{Haw76b}.

A possible resolution of the above paradox is that the radiation from
a black hole is {\em thermal} in the same sense that radiation from
burning coal is thermal, the latter being a consequence of coarse
graining, {\em i.e} averaging over microstates. This requires knowing
what the microstates of a black hole are. 

The microstates should satisfy a number of properties: they should (a)
account for Bekenstein-Hawking entropy, (b) explain Hawking radiation
as averaging over unitary processes and (c) explain why the
time-reversed process of absorption infinitely dominates over emission
in the classical limit. Furthermore, the microstates should (d) be
indistinguishable from one another to the asymptotic observer in the
domain of validity of the no-hair theorems and (e) hopefully give us a
hint regarding the nature of degrees of freedom in gravity which leads
to entropy $\propto$ area.

As is well-known by now, there has been considerable progress in
string theory in recent years on most of the above issues.  In these
lectures, we will mostly focus on the example of the five dimensional
black hole \cite{Str-Vaf96,Cal-Mal96}.  In Sec. 2 we will discuss how
the classical solution is arrived at. In order to have the discussion
reasonably self-contained we will include some techniques of
constructing classical solutions in M/string theory. In Sec. 3 we will
discuss the D-brane picture of the five dimensional black hole.  This
will include the description of microstates in terms of a conformal
field theory and its relation to the world-volume gauge theory of the
D5 branes.  In Sec. 4 we will discuss dynamical questions like
absorption and Hawking radiation. We will first present the
semiclassical calculation and then the derivation in the D-brane
picture. In Sec. 5 we will discuss some open questions. Although we
will use elements of AdS/CFT corrrespondence 
\cite{Mal97,Wit98-ads,Gub-Kle-Pol98} from time to time, we
will not have space here to go into its detailed exposition (see,
e.g., \cite{Aharony} for a review).

\section{\large Construction of the classical solution}

\tgap

The aim of this section will be to construct the classical solution
representing the five-dimensional black hole in \cite{Cal-Mal96}.
Rather than presenting the solution and showing that it solves the low
energy equations of type II superstring, we will describe some aspects
of the art of solution-building.  There are many excellent reviews of
this area (see, for example,
\cite{Ste98,Obe-Pio98,Gau97,Mal96,Cve-review}, other general reviews
on black holes in string/M theories include
\cite{Peet,Skenderis,Duff}), so we shall be brief.  The method of
construction of various classical solutions, we will see, will throw
light on the microscopic configurations corresponding to these
solutions.

Two widely used methods for construction of classical
solutions are

\ni (a) the method of harmonic superposition\nl
(b) $O(d,d)$ transformations
 
\ni We will mainly concentrate on the first one below.

As is well-known by now, classical solutions of type II string
theories can be obtained from those of M-theory
\cite{Tow-Mtheory,Wit-Mtheory} through suitable compactification and
dualities. We will accordingly start with classical solutions of
M-theory, or alternatively, of 11-dimensional supergravity. 

We should note two important points:  

\ni (a)For classical supergravity description of these solutions to be
valid, we need the curvature to be small (in the scale of the
11-dimensional Planck length $l_{11}$ for solutions of M-theory, or of
the string length $l_s$ for string theories)\nl (b) Since various
superstring theories are defined (through perturbation theory) only in
the (respective) weakly coupled regimes, in order to meaningfully talk
about classical solutions of various string theories meaningfully we
need the string coupling also to be small.

For the RR charged type II solutions (charge Q) that we will describe
below, both the above conditions can be met if $Q \gg 1/\gst \gg 1$
(that is, $\gst Q \gg 1, \gst \ll 1$).

\subsection{\large{\bf Classical solutions of M-theory}}

\tgap

The massless modes are the 11-dimensional metric $G_{MN}$,
the gravitino $\psi_M$ and a three-form $A_{MNP}, M=0,1,\ldots,10$. 

The  classical action is
\be
\label{2.1}
S_{11} = \frac{1}{2 \kappa_{11}^2}\int d^{11}x
[ \sqrt{-G} (R - {1\over 48} (dA)^2) - 
\frac{1}{6} A \wedge dA \wedge dA] + {\rm fermions}
\ee
There are two important classical solutions of this Lagrangian, the M2
and M5 branes, whose intersections account for 
most  stable supersymmetric solutions of M-theory
\cite{Pap-Tow96,Tse96,Gau97}.

\sgap

\ni {\bf The 2-brane of M-theory: M2}\cite{M2-ref}

\tgap

\ni We will discuss only this case in some detail. 

\ni Statement of the problem: we want to find (a) a relativistic 2-brane
solution of \eq{2.1} (say stretching along $x^{1,2}$) with (b) 
some number of unbroken supersymmetries.

Condition (a) implies that the solution must have a $SO(2,1)_{0,1,2}
\times$ \nl $ SO(8)_{3,4,5,6,7,8,9,10}$ symmetry, together with
translational symmetries along $x^{0,1,2}$. The subscripts denote
which directions are acted on by the $SO$ groups.

\ni This uniquely leads to
\bea
\label{2.2}
ds_{11}^2 &=& e^{2A_1(r)}  dx^\mu dx_\mu + e^{2A_2(r)} dx^m dx_m
\nn\\
A &=& e^{A_3(r)} dt\wedge dx^1 \wedge dx^2 
\eea
where $\mu=0,1,2$ denote directions parallel to the world-volume
and $m=3,\ldots,9,10$ denote the transverse directions. $r^2
\equiv x^mx_m$.
 
Condition (b) implies that there should 
exist a non-empty set of supersymmetry transformations
$\epsilon$  preserving the solution \eq{2.2}; in particular the gravitino
variation 
\bea
\label{2.3}
\delta_\epsilon \psi_M &=& D_M \epsilon 
+ \frac{1}{288} (\Gamma_M^{NPQR} - 8 \delta_M^N \Gamma^{PQR})
F_{NPQR} \epsilon =0\nn\\
D_M \epsilon &=& 
\left(\del_M + {1\over4} \omega^{BC}_A\Gamma_{BC} \right)\epsilon
\eea
must vanish for some $\epsilon$'s.

It is straightforward to see that Eqn. \eq{2.3} vanishes for $M=\mu$
(world volume directions) if
\bea
\label{2.4}
\del_\mu \epsilon &=& 0, \nn\\
A_3 &=& 3 A_1 
\eea
and
\be
\label{2.5}
\Gamma^{\c{0}\c{1}\c{2}}\epsilon = \epsilon
\ee
where the caret ~~$\c{}$~~ denotes local Lorenz indices. 
(Flipping the sign
of $A$ would correspond to $-\epsilon$ on the
right hand side of \eq{2.5}: this would correspond
to an anti-brane solution in our convention.)

The $M=m$ (transverse) components of \eq{2.3} give rise to 
the further conditions
\bea
\label{2.6}
A_1 &=& - 2 A_2 \nn\\
\epsilon &=& e^{A_3/6} \epsilon_0
\eea

\ni\underbar{Harmonic equation}

The equations \eq{2.4} through \eq{2.6} fix the
three functions $A_i$ in \eq{2.2} in terms of
just one function, say $A_3$. It is easy to
determine it by looking at the equation of motion
of the three-form potential:
\be
\label{2.7}
\del_M(\sqrt{-g} F^{MNPQ}) + {1\over 2.(4!)^2}
\epsilon^{NPQABCDEFGH}F_{ABCD} F_{EFGH}=0
\ee
The second term is clearly zero for our ansatz
\eq{2.2} for $A$. The first term, evaluated for
$(P,Q,R)=(0,1,2)$ gives 
to
\be
\del_m\del_m(e^{-A_3})=0
\ee

\tgap

\ni Thus, the full M2 solution is given by
\bea
\label{2.8}
ds_{11}^2 &=& H^{1/3}[ H^{-1}  dx^\mu dx_\mu +  dx^m dx_m]
\nn\\
A &=&   H^{-1} dt\wedge dx^1 \wedge dx^2
\eea
where $H=H(r)$ satisfies the harmonic equation
in the transverse coordinates
\be
\label{2.9}
\del_m \del_m H = 0
\ee
The simplest solution for $H$,  in an  asymptotically flat space,
is given by 
\be
\label{harmonic-function}
H= 1 + k/r^6
\ee 
Clearly, {\em multi-centred solutions} are also allowed:
\be
\label{multi-centred}
H= 1 + \sum_i {k_i\over | \vec{x} - \vec{x_i}|^6}
\ee
where $\vec x$ denotes the transverse directions $x^m$.

We note that, the constant, $1$, in \eq{harmonic-function} is
essentially an integration constant. Clearly, it can also be zero;
such choices have led to M/string theory solutions involving AdS
spaces.  The point of this remark is to emphasize that the
near-horizon geometry ($r\to 0$), important in the context of AdS/CFT
correspondence \cite{Mal97,Wit98-ads,Gub-Kle-Pol98}, in which $H=
k/r^6$, corresponds to a complete solution in its own right without
the appendage of the asymptotically flat regions. We will return
to the AdS/CFT correspondence several times in these lectures.

\sgap

\ni\underbar{ADM mass}

The integration constant $k$ 
in \eq{harmonic-function} affects the asymptotic fall-off of the
metric as well as of the field strength, and is related
to the ADM mass (per unit area of the 2-brane) $M$ and to the 
gauge charge (per unit area) $q$.
Using the definitions\footnote{We follow
the normalizations in \cite{Ste98} which 
differ from, e.g. \cite{Mal96}.}
\bea
\label{mass}
M &=& \int_{S^7} d^7\Sigma^m (\del^n h_{mn} - \del_m h)
\\
\label{charge}
q &=& \int_{S^7} d^7\Sigma^m F_{m012}
\eea
we get
\be
\label{mass-charge}
M = 6k \Omega_7 = q
\ee
Here $S^7$ represents the sphere at $r^2=x^mx_m=\infty$,%
\footnote{The total
ADM mass, which diverges, includes 
integrals over $x^{1,2}$ as well; we ignore
them here since we are interested in the mass per unit area.
Similar remarks apply to the charge.} 
$h_{MN} \equiv g_{MN} - \eta_{MN} $, $h \equiv \sum_{M=1}^{10} 
h_{MM}$, and $\Omega_n \equiv 2 \pi^{(n+1)/2}/\Gamma({n+1\over 2})$
is the volume of the unit sphere $S^n$.

\sgap

\ni\underbar{BPS solution}

The mass-charge equality in the last equation \eq{mass-charge} is
characteristic of a ``BPS solution''. We provide a very brief 
introduction below. The 11-dimensional supersymmetry algebra 
\cite{Cre-Jul-Sch78} is
\be
\label{susy-algebra}
\{Q, Q\} = C(\Gamma^M P_M + \Gamma^{MN} U_{MN} +
\Gamma^{MNPQR}V_{MNPQR}),
\ee
where $C$ is the charge conjugation matrix and $P,U$ and $V$
are various central terms. When \eq{susy-algebra}
is  evaluated \cite{Azc-Gau-Izq-Tow89} for the above M2 solution, we get 
\be
\label{s-a-m2}
{1\over V_2} \{Q_\alpha, Q_\beta \}= (C \Gamma^\c0)_{\alpha\beta}
M +  (C\Gamma^{\c1\c2})_{\alpha\beta} Q  
\ee
using the notation
\be 
P_\c0 = V_2 M, \quad  U_{\c1\c2} = V_2 q
\ee
where $V_2$ is the spatial volume of the 2-brane (assumed
compactified on a large $T^2$). 

Now, the positivity of the $Q^2$ operator implies that
\be
M \ge q
\ee
where the inequality is saturated when the right hand side
of \eq{s-a-m2} has a zero eigenvector. For our solution
\eq{2.8}, we see from \eq{2.5} 
that the unbroken supersymmetry transformation 
parameter satisfies 
\be
\label{unbroken-susy}
(1 - \Gamma^{\c0\c1\c2})\epsilon =0
\ee
This clearly leads to $M=q$.  This is a typical example of how
classical solutions with (partially) unbroken supersymmetries satisfy
the extremality condition mass= charge.

\sgap

\ni\underbar{``Black brane''} 

The M2-brane itself has a black hole geometry.  If we compactify the
directions $1,2$ on a 2-torus, we have a \underbar{black hole}
solution in the remaining nine extended dimensions.  The compactified
solution is constructed by placing the multiple centres $\vec{x_i}$ in
\eq{multi-centred} at the sites of a lattice defining the 2-torus. The
horizon is situated at $r=0$.  The detailed geometry has been
discussed in many places, {\it e.g} in \cite{Ste98}. Since our main
object of interest is the five-dimensional black hole, and we will use
the M2-brane as essentially a building block for that solution, we
defer the geometrical discussion till we discuss the latter.

Without compactification too, the above solution is ``black'', but it
has extensions in 1,2 directions and is called a black 2-brane.

\sgap

\ni\underbar{Broken supersymmetries reappear
in the near-horizon limit}

The remaining half of the supersymmetry transformations,
leaving the ones in \eq{unbroken-susy}, are 
non-linearly   realized in the M2 geometry and can be regarded
as  spontaneously broken supersymmetries. 
Interestingly, the supersymmetry variations
under these transformations vanish
in the near-horizon limit which has  the geometry
\cite{Gib-Tow93}
\be
AdS_4 \times  S^7
\ee
As a result the broken supersymmetry transformations
reemerge as unbroken, leading to  an enhancement
of the number of supersymmetry charges $16\to 32$
in the near-horizon limit, a fact that plays a crucial
role in the AdS/CFT correspondence. 

\sgap

\ni {\bf Intersecting M2-branes ($M2 \perp M2$)}

\tgap

\ni We will now use the above solution as a building block to
construct more complicated solutions corresponding to intersecting
branes.

We consider first two orthogonal M2 branes, along $x^{1,2}$
and $x^{3,4}$ respectively. The geometry
of the solution corresponds to 
a spacetime symmetry consisting
of rotations  $SO(2)_{1,2}\times SO(2)_{3,4} \times
SO(6)_{5,6,7,8,9,10}$ plus Killing vectors $(\del_t,
\del_1,\ldots, \del_4)$. This leads to
\bea
ds_{11}^2 &=& e^{2A_1} (-dt^2) + e^{2A_2} (dx_1^2 + dx_2^2)
+ e^{2A_3} (dx_3^2 + dx_4^2) + e^{2A_4} dx_i dx_i \nonumber\\
A &=& e^{A_5} dt\wedge dx^1 \wedge dx^2 + 
e^{A_6} dt\wedge dx^3 \wedge dx^4
\eea

\tgap
\hfill  --- o --- \hfill
\tgap

\ni\underbar{Delocalized nature of the solution}

We note that the ansatz above represents a ``delocalized solution''.
A localized $M2 \perp M2$ intersection would destroy translational
symmetries along the spatial world-sheet of both the 2-branes. The
subject of localized intersection is interesting in its own right
(see, e.g. \cite{Sur-Mar98} which is especially relevant to the D1/D5
system), although we do not have space to discuss them here. The
delocalization here involves ``smearing'' the first M2 solution along
the directions 3,4 (by using a continuous superposition in
\eq{multi-centred}, see e.g. \cite{Gau97}), and ``smearing'' the
second M2 solution along 1,2.

\tgap
\hfill  --- o --- \hfill
\tgap

Now, as before, the desire to have a BPS solution leads to
existence of unbroken supersymmetry, or
$\delta_\epsilon \psi_M=0$. This now yields \underbar{four} 
different type of equations, depending on whether
the index $M$ is $0, \{1,2\},\{3,4\}$ or the rest. 
These express the six functions above
in terms of two independent functions $H_1, H_2$.
These functions turn out to harmonic  in the
common transverse directions when one imposes
closure of SUSY algebra or equation of motion. The solution
ultimately is
\bea
ds_{11}^2 &=& (H_1 H_2)^{1/3}
[ -{dt^2\over H_1 H_2 } + 
{dx_1^2 + dx_2^2\over H_1} + {dx_3^2 + dx_4^2\over H_2} 
+ dx_i dx_i]
\nn\\
A &=& {1\over H_1} dt\wedge dx^1 \wedge dx^2 +  
{1\over H_2} dt\wedge dx^3 \wedge dx^4
\eea
The above is an example of ``harmonic superposition of branes''.
(see, e.g., \cite{Tse96}).

\sgap

\ni {\bf M2 $\perp$ M2 $\perp$ M2}

\tgap

\ni Extending the above method, we get the following supergravity solution
for three orthogonal M2-branes, extending respectively along
$x^{1,2}, x^{3,4}$ and $x^{5,6}$:
\bea
ds_{11}^2 &=& (H_1H_2H_3)^{1/3}
[(H_1 H_2 H_3)^{-1} (-dt^2) + 
H_1^{-1} (dx_1^2 + dx_2^2) 
\nn\\
&+& H_2^{-1}  (dx_3^2 + dx_4^2) +
H_3^{-1}  (dx_5^2 + dx_6^2)  +  dx_i dx_i]
\nn\\
A &=& H_1^{-1} dt\wedge dx^1 \wedge dx^2 
+  
H_2^{-1} dt\wedge dx^3 \wedge dx^4
\nn\\
&~~& +
H_3^{-1} dt\wedge dx^5 \wedge dx^6
\eea

\tgap

\subsection{\large \bf The 6D black string solution of IIB on $T^4$}

\tgap

In the following we will construct solutions of type II string
theories using the above M-theory solutions by using various duality
relations which we will describe as we go along. For an early account
of black p-brane solutions in string theory, see \cite{Hor-Str91}.

\tgap

We apply the transformation $ T_{567} R_{10}$   to
the $M2 \perp M2$ solution:

\tgap

\begin{tabular}{l  c  l c l  }
M-theory  & ${\buildrel R_{10} \over \rightarrow}$
        & IIA & ${\buildrel T_{567} \over \rightarrow}$
        & IIB               \\
        &           &           &         &                      \\
M2 (8,9)   &        & D2 ( 8,9) &         & D5 ( 5,6,7,8,9)      \\
M2 (6,7)   &        & D2  (6,7) &         & D1 ( 5)              \\
\end{tabular}

\tgap

The first transformation $R_{10}$ denotes the reduction
from M-theory to type IIA. To do this, one first
needs to compactify the $M2 \perp M2$ solution
along $x^{10}$ (by using the multi-centred harmonic
functions, with centres separated by a distance $2\pi 
R_{10}$
along $x^{10}$). Essentially, at transverse
distances large compared to $R_{10}$, this amounts
to replacement of the harmonic function $1/r^4$ by $1/r^3$
and a suitable modification of the integration constant
to reflect the appropriate quantization conditions.
At this stage, one still has 11-dimensional fields. To
get to IIA fields, we use the reduction formula
\bea
ds_{11}^2 &=& \exp[-2\phi/3] ds_{10}^2 + \exp[4\phi/3] (dx^{10}
        + C^{(1)}_\mu dx^\mu )^2
\nn\\
A &=& B \wedge dx^{10} + C^{(3)}
\eea
It is instructive to verify at this stage that the classical D2
solutions do come out of the M2-brane after these transformations. We
use the notation $C^{(n)}$ for the $n$-Form Ramond-Ramond (RR)
potentials in type II theories.

The second transformation $T_{567}$ involves a sequence of T-dualities
(for a recent account of T-duality transformations involving RR
fields, see \cite{Has99}). We denote by $T_m$ T-duality along the
direction $x^m$.  $T_{567}$ denotes $T_5 T_6 T_7$.

The final transformation, not explicitly written in the above table,
is to wrap $x^{6,7,8,9}$ on $T^4$. We will
denote the volume of the $T^4$ by
\be
V_{T^4} \equiv \alpha^{\prime 2} (2\pi)^4 \tilde v
\ee

Assuming the number of the two orthogonal sets of M2-branes to be
$Q_5, Q_1$ respectively, the final result is: $Q_5$ strings from
wrapping D5 on $T^4$ and $Q_1$ D-strings. This is the
D1/D5 system in IIB supergravity, characterized by the
following solution:
\bea
\label{6d-black-string}
ds_{10}^2 &=& f_1^{-1/2}f_5^{-1/2} (-dt^2 + dx_5^2) 
 +  f_1^{1/2}f_5^{1/2} dx_i dx^i  +
                f_1^{1/2}f_5^{-1/2} dx_a dx^a
\nn\\
f_{1,5} &=& (1 + r^2_{1,5}/r^2)  
\nn\\
r_1^2 &=& \gst Q_1 \alpha'/\tilde v,\quad  r_5^2 = \gst Q_5 \alpha'
\nn\\
B'_{05} &=& -\half (f_1^{-1} -1)  \nn\\
dB'_{ijk} &=& \epsilon_{ijkl} \del_l f_5 \nn\\
e^{-2\phi} &=& f_5 f_1^{-1}
\eea
Here $B'\equiv C^{(2)}$, the 2-form RR gauge potential
of type IIB string theory.

\tgap

\subsection{\large\bf The extremal 5D black hole solution}

\tgap

Let us now compactify $x^5$ along a circle of radius $R_5$ and wrap
the above solution along $x^5$ to get a 0-brane in five dimensions.
Let us also ``add'' gravitational waves (denoted $W$) moving to the
left along $x^5$. This gives us the BPS version
\cite{Str-Vaf96,Cal-Mal96} of the five-dimensional black hole. Adding
such a wave can be achieved by augmenting the $M2\perp M2$ solution by
a third, transverse, set of M2-branes along $x^{5,10}$ and passing
through the same sequence of transformations as above.  Alternatively,
one can apply the Garfinkle-Vachaspati transformation \cite{Gar-Vac92}
to the above solution, with the same final result.  In the first
method, one should start from the $M2\perp M2\perp M2$ solution and
use a transformation table like the above which has an additional line

\tgap

\begin{tabular}{l  c  l c  l }
M2 ( 5,10)  & $\to$       & NS1 ( 5 )  &  $\to$       & W  ( 5)   \\
\end{tabular}

\tgap

\ni The last transformation essentially reflects the fact
that T-duality changes winding modes to momentum modes. 
($W$ denotes a gravitational \underbar{w}ave and  not
a \underbar{w}inding mode.)

The final configuration corresponds to D5 branes along
$x^{5,6,7,8,9}$ and D1 branes along $x^5$, with 
a non-zero amount of
(left-moving) momentum. If the number of the three sets of M2 branes
are $Q_1,Q_5$ and $N$ respectively, then these will correspond to the
numbers of D1-, D5-branes and the quantized left-moving momentum
respectively.

The final solution is given by
\bea
\label{extremal-5d}
ds_{10}^2 &=& f_1^{-1/2}f_5^{-1/2} (-du dv + (f_n-1) du^2) 
\nn\\
        &~~& +  f_1^{1/2}f_5^{1/2} dx_i dx^i  +
                f_1^{1/2}f_5^{-1/2} dx_a dx^a
\nn\\
f_{1,5,n} &=& (1 + r^2_{1,5,n}/r^2)  
\nn\\
B'_{05} &=& -\half (f_1^{-1} -1)  \nn\\
dB'_{ijk} &=& \epsilon_{ijkl} \del_l f_5 \nn\\
e^{-2\phi} &=& f_5 f_1^{-1}
\eea

The \underbar{spacetime symmetry} $\S$ of the above 
solution is:
\be
\label{space-time-symmetry}
\S= SO(1,1) \times  SO(4)_E \times `SO(4)_I^,
\ee
where $SO(1,1)$ refers to directions $0,5$, $SO(4)_E$
to directions $1,2,3,4$ ($E$ for external)  $`SO(4)_I$' to
directions $6,7,8,9$ ($I$ stands for internal; the
quotes signify that the symmetry is broken by wrapping
the directions on a four-torus although for low energies
compared to the inverse radii it remains a symmetry
of the supergravity solution).

\tgap

The unbroken \underbar{supersymmetry} can be read off either
by recalling those of the M-theory solution and following the dualities 
or by solving the Killing spinor equations (analogous to  \eq{2.3}). 
The result is: 
\bea 
\Gamma^{056789} \epsilon_L &=& \epsilon_R
\nn\\
\Gamma^{05} \epsilon_L &=& \epsilon_R
\nn\\
\label{supersymmetry}
\Gamma^{05} \epsilon_{L,R} &=& \epsilon_{L,R}
\eea
The first line corresponds to the unbroken supersymmetry appropriate for
the D5-brane (extending in 5,6,7,8,9 directions). The second line
refers to the D1-brane.  The last line corresponds to unbroken
supersymmetries in the presence of the gravitational wave.
(The superscripts in $\Gamma^{ab..}$ denote local
Lorenz indices like in \eq{2.3}, although we have dropped 
the carets.)

The parameters $r_{1,5,n}^2$ in
\eq{extremal-5d} are related to the
integer-quantized charges $Q_{1,5}$ and momentum $N$ by
\bea 
\label{quantization}
r_{1,5}^2 &=&  c_{1,5} Q_{1,5}
\nn\\
r_n^2 &=&  c_n N
\eea
where
\bea
\label{coefficients}
c_1 &=& {4G^5_N R_5\over \pi \alpha'}
\nn\\
c_5 &=& \gst \alpha' \nn\\
c_n &=& {4G_N^5 \over \pi R_5} \nn\\
G_N^5 &=& G_N^{10}/ (2\pi R_5 V_{T^4}),\quad G_N^{10}=8\pi^6\gst^2
\alpha^{\prime 4} 
\eea
For a detailed discussion of quantization conditions
like \eq{quantization},
see, e.g. \cite{Bre-Lu-Pop-Ste97,Mal96,Ste98}. Here
$G_N^d$ denotes the $d$-dimensional Newton's constant.

We defer the discussion of the geometry and the Bekenstein-Hawking
entropy till the next section where we describe the non-extremal
version.

\tgap

\subsection{\large\bf Non-extremal five-dimensional black hole}

\tgap

We have explained above how to construct from first principles the BPS
(hence extremal) version of the 5D black hole solution. We will now
present an algorithm (without proof and specialized to intersections
of M2) of how to generalize these constructions to their non-extremal
(nonsupersymmetric) versions \cite{Cve-Tse96}:

\sgap

\ni\underbar{Rule 1}: 
In the  transverse part of the metric (including time) make
the following substitution:
\bea 
\label{rule1}
dt^2 &\to& h(r) dt^2,\quad  dx^i dx_i \to h^{-1}(r) dr^2 + r^2 
d\Omega_{d-1}^2
\nn\\ 
h(r) &=& 1 - \mu/r^{d-2}\nn
\eea
with the harmonic function now defined as
\be 
H(r) = 1 + \tilde Q/r^{d-2} 
\ee
where $\tilde Q$ is a combination
of the non-extremality parameter $\mu$ and some ``boost''
angles:  
\be 
\label{boost}
\tilde Q = \mu \sinh^2 \delta
\ee
(for multicentred solutions, 
$ \tilde Q_i = \mu \sinh^2 \delta_i$ etc.)

\sgap

\ni\underbar{Rule 2}: 
In the expression for $F_4= dA$, make the substitution
\bea
H &\to& \tilde H(r)= 1+ \frac{\bar Q}{r^{d-2} + \tilde Q
- \bar Q} = \left( 
1- \frac{\bar Q}{r^{d-2}}H^{-1}
\right)^{-1},
\nn\\ 
\bar Q &=& \mu \sinh \delta \cosh \delta 
\eea

\sgap

\ni\underbar{A heuristic motivation for the algorithm}

We present a brief, heuristic, motivation for the
above algorithm. Suppose we view a static Schwarzschild
black hole, of ADM mass $m$, from the five-dimensional Kaluza-Klein
viewpoint. The 5-momenta $(p_0, \vec p, p_5)$ will
be given by
\be
p_0 = m, \quad  p_5 \propto \hbox{charge} =0, 
\quad  \vec p=0
\ee
The second equation follows because the Schwarzschild 
black hole is neutral.

A way of generating charged solutions is to consider
the above solution in five non-compact dimensions
and perform a boost in the 0-5 plane. The momenta transform as
\be
p'_0 \equiv M = m \cosh \delta, 
\quad p'_5 \equiv \tilde Q 
= m \sinh \delta, \quad  {\vec p}'=0
\ee 
We can now wrap the fifth dimension to get a
charged black hole in four non-compact dimensions.
The extremal limit ($M=Q$) can be attained by
\be
\delta\to \infty, m\to 0, m e^\delta \to {\rm constant}
\ee
so that  
\be
\label{approach-extremal}
\tilde Q \to M = me^\delta/2, \quad  p'_R \equiv p'_0 - p'_5\to 0 
\ee
\ni{\sl Near-extremal limit}\nl
The near-extremal limit is obtained by 
keeping the leading corrections in $e^{-\delta}$. Thus,
\be
\tilde Q/M = \tanh\,\delta \simeq 1 - m^2/(2 
{\tilde Q}^2),\qquad  p'_R \ll p'_L
\ee
In terms of these parameters, the four-dimensional metric for a 
near-extremal charged (RN) black hole is given by
\bea
ds_4^2 &=& - f dt^2 + f^{-1} dr^2
+ r^2 d\Omega_2^2 \nn\\
f &\equiv& 1 - 2M/r + {\tilde Q}^2/r^2 = f_{\rm ext} h(r)
\nn\\
f_{\rm ext} &\equiv &  (1 - {\tilde Q}/r)^2,  
\quad  h(r)= (1 - \mu/r) 
\nn\\
\mu &=& m^2/\tilde Q  
\eea
The last equality implies
\be
\tilde Q= \mu \sinh^2 \delta,
\ee
same as \eq{boost} above. Also, the second equality
agrees with Rule 1 for relating the non-extremal $g_{tt},
g_{rr}$ to their extremal counterparts. 

Of course, we have considered here only the near-extremal case. The
remarkable thing about the algorithm mentioned above is that it works
for arbitrary deviations from extremality.

\sgap

\ni Applying this rule to the $M2\perp M2 \perp M2$ case, 
we get 
\bea
ds_{11}^2 &=& (H_1 H_2 H_3)^{-1/3} [ - H_1 H_2 H_3 h dt^2 +
H_1 (dy_1^2 + dy_2^2)  
\nn\\
&+& H_2 (dy_3^2 + dy_4^2) +  H_3
 (dy_5^2 + dy_6^2) + h^{-1} dr^2 
\nn\\ 
&+& r^2 d\Omega_{d-1}]
\eea

The rest of the story is similar to the BPS case. We first reduce the
M-theory solution to IIA, wrap the solution on a $T^5$ and then
T-dualize to IIB.

Under the reduction from M-theory to type IIA in ten dimensions,
we get
\bea
e^{-2\phi} &=& F_1 F_5^{-1}
\nn\\
ds_{10}^2 &=& F_1^{-1/2} F_5^{-1/2}[-dt^2 + dx_5^2 
\nn\\
&+& (1-h)(\cosh \alpha_n dt + \sinh \alpha_n dx_5)^2]
\nn\\
&+&  F_1^{1/2}F_5^{1/2}
(\frac{dr^2}{h} + r^2 d\Omega_3^2)
\nn\\
&+& F_1^{1/2}F_5^{-1/2} dx_a dx^a 
\eea
where $a=6,..,9$, $(r, \Omega_3)$ are polar coordinates
for $x^{1,2,3,4}$ and 
\bea
F_s &=& 1 + \frac{r_s^2}{r_0^2},\, r_s^2 = r_0^2 \sinh^2 \alpha_s,
\, s=1,5
\nn\\
h &=& (1 - \frac{r_0^2}{r^2}) 
\eea
The parameter $r_0^2$ is  the same as the non-extremality parameter
$\mu$ of \eq{rule1}, while $\alpha_{1,5,n}$ are related
to the boost angle of \eq{boost}.

It is easy to compactify the directions $x^{5,6,7,8,9}$ on a
$T^5$ (radii $R_{5,6,7,8,9}$, $V\equiv R_6R_7R_8R_9$), 
using the reduction formula
\be
ds_{10}^2 = e^{2\chi} dx_a dx^a e^{2\psi}(dx_5 + A_\mu dx^\mu)^2
+ e^{-(8\chi + 2\psi+\phi)/3} ds_5^2
\ee
(the first two exponential factors are simply the definitions of the
scalars $\chi, \psi$; the factor in front of $ds_5^2$ can be found
easily by demanding that $ds_5^2$ is 
the five-dimensional Einstein metric). Here $\mu=1,2,3,4$. 

This is still a IIA solution. In order to get the IIB version,
we have to apply  the sequence $T_{567}$. We omit the
details here which are fairly straightforward. At the
end we  get the following five-dimensional Einstein metric
\cite{Hor-Mal-Str96}:
\bea
\label{non-extremal-5d}
ds_5^2 &=& - h f^{-2/3} dt^2 + f^{1/3}(\frac{dr^2}{h} + 
r^2 d\Omega_3^2)
\nn\\ 
h &=& 1 - r_0^2/r^2
\nn\\
f &=&  F_1 F_5 (1 + r_n^2/r^2) 
\nn\\
r_n^2 &=& r_0^2 \sinh^2 \alpha_n 
\eea
There are six independent parameters of the metric 
$\alpha_{1,5,n}$,
$r_0, R_5,\tilde v \equiv V_{T^4}/(2\pi l_s)^4$
($l_s = \sqrt{\alpha'}$). 
The boost angles and the non-extremality
parameters   are related to the three charges and the mass
$M$  as follows: ($H\equiv dB'$)
\bea
\label{charges}
Q_1 &=&\frac{V}{4\pi^2g}\int e^{\phi} * H=  
\frac{\tilde v r_0^2}{2 \alpha'\gst} \sinh 2\alpha_1
\nn\\
Q_5 &=&\frac{1}{4\pi^2 \gst}\int  H= 
\frac{r_0^2}{2 \alpha \gst}\sinh 2\alpha_5
\nn\\
N  &=& \frac{R_5^2 \tilde v r_0^2}{2\alpha\gst^2} \sinh 2 \alpha_n
\nn\\
M &=& \frac{R_9 \tilde v r_0^2}{2 l_s^3 \gst^2}
(\cosh 2\alpha_1 + \cosh 2\alpha_5+ \cosh 2\alpha_n)
\eea
There is another very interesting representation of the above-mentioned
six parameters in terms of what appears as brane-, antibrane-numbers
and left-,right-moving momenta:
\bea
N_{1,\bar 1} &=& \frac{\tilde v r_0^2}{4\alpha'\gst} e^{\pm 2\alpha_1}
\nn\\
N_{5,\bar 5} &=& \frac{r_0^2}{4\alpha'\gst}  e^{\pm 2\alpha_5}
\nn\\
N_{L,R}  &=& \frac{R_5^2 \tilde v r_0^2}
{4\alpha^{\prime 2}\gst^2} e^{\pm 2 \alpha_n}
\eea
Clearly $N_1 - N_{\bar 1}= Q_1, N_5 - N_{\bar 5}= Q_5$
and $N_L - N_R= N$. The extremal limit corresponds to
taking $r_0\to 0, \alpha_i \to \infty$ keeping the charges
$Q_{1,5}, N$ finite.

\sgap

\ni\underbar{Geometry}

It is easy to see that the above solution is a five-dimensional
black hole, with horizon at $r=r_0$. The horizon has a finite
area $A_h$, given by
\bea
\label{area}
A_h &=& 2\pi^2 r_0^3 \cosh\alpha_1 \cosh\alpha_5\cosh\alpha_n
\nn\\  
&=& 8 \pi G_N^{5} (\sqrt N_1
+\sqrt N_{\bar 1})(\sqrt N_5
+\sqrt N_{\bar 5})(\sqrt N_L+ \sqrt N_R)
\eea
Here we have used \eq{charges} and the value of
the five-dimensional Newton's constant $G_N^5$
in \eq{coefficients}. 

\begin{figure}[!ht]
\begin{center}
\leavevmode
\epsfbox{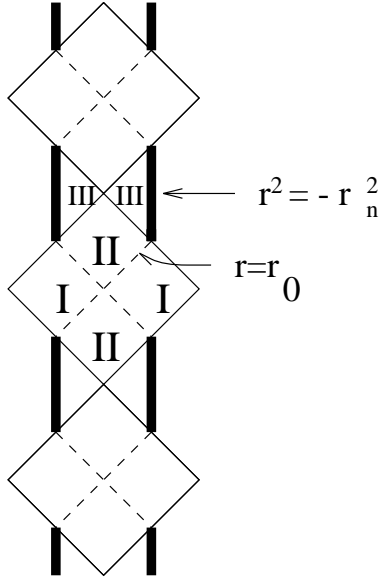}
\end{center}
\caption{Carter-Penrose diagram for the non-extremal 5D black hole}
\label{f-penrose}
\end{figure}

The fact that the horizon has a finite area indicates that
the singularity lies ``inside'' $r=r_0$. It is not
at $r=0$, however, which corresponds to the inner horizon
(where light-cones ``flip'' the second time as one
travels in). To locate the singularity one needs
to use other coordinate patches which extend the manifold
further ``inside''. The singularity is time-like
and the Carter-Penrose diagram (\fig{f-penrose}) %
is similar to that of the non-extremal Reissner-Nordstrom metric.

\sgap

\ni\underbar{Bekenstein-Hawking entropy}

By using the formula $S= {1\over 4 G_N^5} A_h$, we get
\be
\label{non-extremal-entropy}
S_{\rm BH} = 2\pi (\sqrt N_1
+\sqrt N_{\bar 1})(\sqrt N_5
+\sqrt N_{\bar 5})(\sqrt N_L+ \sqrt N_R)
\ee
The extremal entropy is given by
\be
\label{extremal-entropy}
S_{\rm BH} = 2\pi \sqrt{Q_1 Q_5 N}
\ee
Both the above formulae are U-duality invariant,
in the following sense.
Consider an $S(3)$ subgroup
of the U-duality group of type IIB on $T^5$, which permutes
the three charges $Q_1, Q_5$ and $N$. Such an $S(3)$ is
generated by
(a) $T_{6789}$  which sends $Q_1\to Q_5, Q_5\to Q_1,
N\to N$, and \\
(b) $T_{9876}ST_{65}$ which sends $Q_1 \to Q_5,
Q_5\to N, N\to Q_1$.\\
The entropy
formula \eq{extremal-entropy}
remains invariant under these permutations. Since the
``anti''-objects are also permuted among each other
by these U-duality transformations, we can say that
the entropy formula \eq{non-extremal-entropy} is
also U-duality invariant.

\tgap

\section{\large D-brane picture}

\sgap

D-branes are solitonic configurations of superstring theories
which, by definition, are characterized by low energy excitations that
are open strings ending on them. For a review, see, e.g.
\cite{Pol-tasi96}. We will provide a very brief introduction
here.

\sgap

\ni {\bf Dp branes}

\sgap

\ni We will discuss a single Dp-brane first.

Open string excitations of a single Dp-brane, extending
in directions $x^{1,2,..,p}$, obey the boundary conditions
\bea
x^M(z) &=& R^M_N x^N(\z)
\nn\\
\psi^M(z) &=& R^M_N \psi^N(\z)
\nn\\
R &=& {\rm diag}[1,1,..,1,-1,-1,..,-1]
\nn\\
S^\alpha (z) &=& {\cal R}^\alpha_\beta S^\beta(\z)
\nn\\
{\cal R} &=& \Gamma^{01...p}
\eea
at $z=\z$. The coordinates $z, \z$ refer to the upper
half plane which is related by a conformal transformation to the disc
geometry of tree-level open-string world-sheet (see, e.g., 
\cite{Has-Kle96} for more details). 
$M$ and $\alpha$ are vector and spinor indices in ten dimensions.
The positive eigenvalues of the matrix $R$ correspond to
the longitudinal directions
$x^\mu, \mu=0,1,..,p$ (Neumann boundary condition) and the negative
eigenvalues correspond to the transverse directions $x^i, i=p+1,..,9$
(Dirichlet boundary conditions).

\underbar{Spacetime symmetry:}\\ 
Clearly these boundary conditions reflect a
$SO(p,1) \times SO(d)$ symmetry as well as 
translational symmetries along $x^{0,1,...,p}$.

\underbar{Supersymmetry:}\\ 
The boundary condition for the spin fields 
implies that the supersymmetry transformation parameters
must satisfy
\be
\epsilon_L= \Gamma^{01...p} \epsilon_R
\ee
We see that only half of the supersymmetries are
preserved, implying the $1/2$-BPS nature
of D$p$-branes. It is easy to see the correspondence to the
classical brane solutions described earlier, both
in terms of the spacetime symmetry and supersymmetry
({\em cf.} Eqns.  \eq{space-time-symmetry},\eq{supersymmetry}).

\sgap

\ni\underbar{Multiple coincident D$p$ branes}:

Spacetime symmetries and supersymmetries remain the same as for the
single brane.  The main additional ingredient is that open strings can
now begin and end on different branes. This attaches an additional
degree of freedom to the end-points of open strings, which can be
identified with {\em Chan-Paton factors}.  Thus, an open string
beginning on the i-th brane and ending on the j-th brane gets labelled
by a matrix $\Lambda^{(ij)}$ whose $(i,j)$ element is 1 and rest are
zero. These matrices generate the $U(N)$ algebra
\cite{Pol-tasi96}. The massless
open string excitations correspond to a supersymmetric $U(N)$
Yang-Mills theory in $p+1$ dimensions, which can be regarded
as a dimensional reduction of ${\cal N}=1$ SYM in 10 dimensions
\cite{Wit-SYM}.

Since the supergravity fields cannot couple to non-singlets
of the SYM theory, the signature of multiple branes is only
in the ADM mass and the total RR-charge in \eq{mass} and \eq{charge}.  

\tgap

\subsection{\large\bf
Microstates corresponding to the five-dimensional black hole}

In  section 2 we have presented the classical solution of the
five-dimensional black hole. The construction of the extremal hole
suggests that we should look for a bound state of $Q_1$ D1 branes and
$Q_5$ D5 branes with total left-moving momentum equal to $N$ (in units
of $1/R_5$).

In order to find the low energy excitations of such a system, we begin
with the D1/D5 system (which corresponds to the 6 dimensional black
string) in which $x^5$ is still non-compact and there is no momentum
mode yet. The low energy excitations correspond to open strings which
can begin on a D1-brane or a D5-brane and end on a D1-brane or a
D5-brane. We call such open strings (1,1), (1,5), (5,1) or (5,5) where
$(p,q)$ denotes an open string beginning on a D$p$- and ending on a
D$q$-brane.

\tgap

\ni The (1,1) strings obey the following boundary conditions: 

\ni DD boundary conditions along directions 1,2,3,4,6,7,8,9\\
NN boundary conditions along directions 0,5 

\ni (By DD we mean that the open strings satisfy Dirichlet
boundary conditions at both ends. The notations ND and NN
are likewise defined)

\tgap

\ni For the (5,5) strings the boundary conditions are:

\ni DD: 1,2,3,4\\
NN: 0,5,6,7,8,9

\tgap

\ni For the (1,5) (and (5,1)) strings the boundary conditions are:

\ni DD: 1,2,3,4\\
NN: 0,5\\
ND: 6,7,8,9

\tgap

It is clear that the open string boundary conditions correspond to the
same spacetime symmetry $SO(1,1) \times SO(4)_E \times `SO(4)_I$' as
in \eq{space-time-symmetry} which characterizes the classical
solution.  It is also easy to see that these boundary conditions
lead to  the same supersymmetries as in \eq{supersymmetry}.

\sgap

\ni{\bf Massless modes}

\tgap

In order to see the massless degrees of freedom
\cite{Mal96,Mal97a,Has-Wad97b,Dav-thesis} let us look at
the following table of zero-point energies (i.e., $L_0$ values for the
Fock space vacua of the oscillators $x, \psi_R, \psi_{NS}$). The table
can be constructed simply from the moding of various fields in their
normal mode expansion for the appropriate boundary condition.

\sgap

\ni\underbar{Table of zero-point energies}

\sgap

\begin{tabular}{ || l || l || l || l || }
\hline
 & NN & DD & ND \\
\hline
X  & $-$1/24 (P) & $-$1/24  (P) & 1/48 (AP) \\
\hline
$\psi_R$ & 1/24 (P) & 1/24 (P) & $-$1/48 (AP) \\
\hline
$\psi_{NS}$ & $-$1/48 (AP) & $-$1/48  (AP) & 1/24 (P) \\
\hline
\end{tabular}

(P=periodic, AP=anti-periodic)

\sgap

\ni 
(1,1) strings: In the light-cone gauge (where we choose $x^0, x^5$,
both non-compact, as the light cone directions), there are 8 DD
directions, leading to a total zero-point energy $L_0 =-1/2$ in the NS
sector and $L_0=0$ in the R sector. We assume that the radii in the
$6,7,8,9$ directions are of the order of the string length; so at low
energies we can ignore any winding modes in these directions.  The
massless spectrum of (1,1) strings, then, is that of a supersymmetric
$U(Q_1)$ gauge theory in 1+1 dimensions, as mentioned above in the
context of a single set of Dp-branes.  The field content can be
organized into the vector multiplet and the hypermultiplet of an
${\cal N}=2$ theory in four-dimensions (or those of $\N=1$ in six
dimensions $x^{0,1,2,3,4,5}$)

\bea 
\mbox{Vector multiplet}&:& \;
A_0^{(1)}, A_5^{(1)}, 
\phi_1^{(1)}, \phi_2^{(1)}, \phi_3^{(1)},
\phi_4^{(1)} \nn\\ 
\mbox{Hypermultiplet}&:& \; Y_6^{(1)},
Y_7^{(1)}, Y_8^{(1)}, Y_9^{(1)} 
\eea 

The $A_0^{(1)}, A_5^{(1)}$ are the $U(Q_1)$ gauge fields in the
non-compact directions. The $Y^{(1)}$'s and $\phi^{(1)}$'s are gauge
fields in the compact directions of the $\N=1$ super Yang-Mills in
ten-dimensions.  All the fields transform as adjoints of $U(Q_1)$. The
hypermultiplet of $\N=2$ supersymmetry are doublets of the $SU(2)_R$
symmetry of the theory.  The $Y^{(1)}$'s can be arranged as doublets
under $SU(2)_R$ as

\be N_{a\bar{a}}^{(1)} = \left(
\begin{array}{c}
N_{1(a\bar{a})}^{(1)} \\ \nonumber
N_{2(a\bar{a})}^{(1)\dagger}
\end{array}
\right)
=\left(
\begin{array}{c}
Y_{9 (a\bar{a})}^{(1)} + i 
Y_{8 (a\bar{a})}^{(1)}  \\
Y_{7 (a\bar{a})}^{(1)} - i 
Y_{6 (a\bar{a})}^{(1)} 
\end{array}
\right)
\ee
where $a, \bar{a}$ runs from $1,\ldots ,Q_1$. 
\\

\sgap

\ni 
(5,5) strings: These 4 DD and 4 NN directions. The massless spectrum
can be found again from the table of zero-point energies. Ignoring the
momentum modes along the $T^4$, we again have a 
$U(Q_5)$ theory in 1+1 dimensions. The field
content is exactly similar to those of the (1,1) strings:

\bea 
\mbox{Vector multiplet}&:& \;
A_0^{(5)}, A_5^{(5)}, 
\phi_1^{(5)}, \phi_2^{(5)}, \phi_3^{(5)},
\phi_4^{(5)} \nn\\ 
\mbox{Hypermultiplet}&:& \; Y_6^{(5)},
Y_7^{(5)}, Y_8^{(5)}, Y_9^{(5)} 
\eea 

The superscript indicates that the fields correspond to
the (5,5) strings and transform as adjoint of $U(Q_5)$.

\sgap

\ni
(1,5) and (5,1) strings: These have 4 ND directions (we do not
differentiate between ND and DN here) and 4 DD directions.  The zero
point energies {\em vanish} in both the NS and R sectors!  From the
fact that $\psi_{NS}$ are periodic for the ND directions, one sees
that the massless mode is a boson transforming as a spinor of
$SO(4)_I$. This gives four  bosons.  The GSO projection
projects out half of these which reduces the number of bosons to $2$.
The two bosons of the $(1,5)$ strings and the $(5,1)$ strings combine
to form a complex doublet transforming under the diagonal $SU(2)$ of
the $SO(4)_I$.  As the hypermultiplets of $\N=2$ theory transform as
doublets under $SU(2)_R$, the diagonal $SU(2)$ of $SO(4)_I$ can be
identified with the $SU(2)_R$ of the gauge theory.  The Chan-Paton
factors show that they transform as bi-fundamentals
$(Q_1, \bar{Q_5})$  of $U(Q_1)\times U(Q_5)$.  
We arrange these hypermultiplets as doublets of
the $SU(2)_R$ symmetry of the theory in the form
\be
\chi_{a\bar{b}} = 
\left(
\begin{array}{c}
A_{a\bar{b}} \\
B^{\dagger}_{a\bar{b}}
\end{array}
\right)
\ee
We note that the fermionic superpartners of these
hypermultiplets which arise from the Ramond sector of the massless
excitations of $(1,5)$ and $(5,1)$ strings carry spinorial indices
under $SO(4)_E$ and they are singlets under $SO(4)_I$.

The gauge theory of the D1/D5 system, therefore, is a $1+1$
dimensional $(4,4)$ supersymmetric gauge theory with gauge group
$U(Q_1)\times U(Q_5)$. The matter content of this theory consists of
hypermultiplets $Y^{(1)}$'s, $Y^{(5)}$'s transforming as adjoints of
$U(Q_1)$ and $U(Q_5)$ respectively. It also has the hypermultiplets
$\chi$'s which transform as bi-fundamentals of $U(Q_1)\times
{U(Q_5)}$.

\sgap

\ni\underbar{Counting of degrees of freedom:}

We will now show that this gauge theory has the required degrees of
freedom to describe the entropy of the extremal D1/D5 black hole. The
D1/D5 bound state is described by the Higgs branch of this gauge
theory. The Higgs branch is obtained by giving expectation values to
the hypers. This makes the vector multiplets massive. For a
supersymmetric vacuum the hypers take values over the surface which is
given by setting the superpotential of the gauge theory to zero.
Setting the superpotential to zero imposes two sets of D-flatness
conditions corresponding to each of the gauge groups $U(Q_1)$ and
$U(Q_5)$. The D-terms for the gauge group $U(Q_1)$ are given by
\cite{Has-Wad97b,Alv-Has-review,Wes-book}\footnote{%
We ignore here Fayet-Iliopoulos terms and the related issue
of singularity at the origin of the Higgs branch (see,
{\em e.g.} \cite{Sei-Wit99,Sei-Wit99b,Dha-Man-Wad-Yog99}).}
\bea
\label{dterm-1}
A_{a\bar{b}}A^{*}_{a'\bar{b}} -B_{b\bar{a}'}B^{*}_{b\bar{a}} 
+ [N_1^{(1)}, N_{1}^{\dagger}]_{a\bar{a}'}
- [N_2^{(1)}, N_2^{(1)}]_{a\bar{a}'} =0 \\ \nonumber
A_{a\bar{b}}B_{b\bar{a}'} + [N_1^{(1)},
N_2^{(1)\dagger}]_{a\bar{a}'} = 0 \\ \nonumber
\eea
while the D-terms of the gauge group $U(Q_5)$ are
\bea
\label{dterm-5}
A_{a\bar{b}'}A^{*}_{a\bar{b}} -B_{b\bar{a}}B^{*}_{b'\bar{a}} 
+ [N_1^{(5)}, N_{5}^{\dagger}]_{b\bar{b}'}
- [N_2^{(5)}, N_2^{(5)}]_{b\bar{b}'} =0 \\ \nonumber
A_{a\bar{b}'}B_{b\bar{a}} + [N_1^{(5)},
N_2^{(5)\dagger}}_{b\bar{b}'] = 0 
\eea
Here $a,a'$ run from $1,\ldots ,Q_1$ and 
$b,b'$ run from $1, \ldots ,
Q_5$.

The total number of bosonic degrees of 
freedom from all the hypermultiplets  ($Y^{(1)}_{a\bar a}
Y^{(5)}_{b\bar b}, A_{a\bar b}, B_{\bar a b}$) is
\be
4Q_1^2 + 4Q_5^2 + 4Q_1Q_5
\ee
The first equation in \eq{dterm-1} is real while the second
equation in \eq{dterm-1} is complex. The total number of
constraints imposed by \eq{dterm-1} is $3Q_1^2$. Similarly the
set of D-term equations in \eq{dterm-5} imposes $3Q_5^2$
constraints. Equations \eq{dterm-1} and \eq{dterm-5} have
the same trace parts corresponding to the vanishing of $U(1)$
D-terms, namely,
\bea
A_{a\bar{b}}A^{*}_{a\bar{b}} -B_{b\bar{a}}B^{*}_{b\bar{a}} &=& 0 \\
\nonumber
A_{a\bar{b}}B_{a\bar{b}} &=& 0
\eea
which are three real equations. Therefore, the vanishing of D-terms
imposes $3Q_1^2 + 3Q_5^2 -3$ constraints on the fields. 
One can use the gauge symmetry $U(Q_1)$ and $U(Q_5)$ to
remove another $Q_1^2 + Q_5^2 -1$ degrees of freedom. The
$-1$ reflects the fact that all the
hypermultiplets are invariant under the diagonal $U(1)$ of
$U(Q_1)\times U(Q_5)$. After gauge fixing, the number of gauge
invariant bosonic degrees of freedom to parameterize the moduli space
is $4(Q_1Q_5 +1)$. 

We are interested in low energy black hole processes so it is
sufficient to study the SCFT of the Higgs branch. The SCFT will have
${\cal N}= (4,4)$ SUSY with central charge $6(Q_1Q_5+1)$ on some
target space ${\cal M}$. (In general, the number of ``degrees of
freedom'' need not simply translate to central charge  owing
to interactions; however, with the extent of supersymmetry present
here, the manifold ${\cal M}$ must be hyperKahler for which the above
claim for the central charge is true.)  To find the microstates
corresponding to the extremal D1/D5 black hole we look for states with
$L_0 =N$ and $\bar{L}_0 =0$.  The asymptotic number of distinct
states of this SCFT given by Cardy's formula \cite{Cardy}
\be
\label{cardy-formula}
\Omega = \exp(2\pi \sqrt{cL_0/6}=
\exp (2\pi \sqrt{Q_1Q_5 N} )
\ee
From the Boltzmann formula one obtains
\be
S=2\pi \sqrt{Q_1Q_5 N}
\ee
This exactly reproduces the Bekenstein-Hawking entropy
\eq{extremal-entropy} of the extremal D1/D5 black hole.
We will remark about the non-extremal black hole shortly.

\subsection{\large\bf Instanton moduli space}

\tgap

We found above that the Higgs branch of the 
gauge theory of the D1/D5 system flows 
in the infrared to  an ${\cal N} =(4,4)$ SCFT 
on a target space ${\cal M}$ with central charge $6Q_1Q_5$. 
For black hole processes like Hawking radiation it is important to
know the target space ${\cal M}$. 
In this section we review the arguments which show that the
target space ${\cal M}$ is a resolution of the orbifold
$(\tilde{T}^4)^{Q_1 Q_5}/S(Q_1Q_5)$. ($\tilde{T}^4$ can 
be different from the 
compactification torus $T^4$.)

We first note that the
world-volume theory of $Q_5$ coincident D5 branes is a  5+1
dimensional U($Q_5$) SYM theory. One way to 
understand D1 branes bound to these D5-branes is
to represent the D1-branes as solitons \cite{Dou95}
of this
SYM theory. The simplest way to see this 
is to note the Chern-Simons coupling in the world-volume 
action of a D5 brane \cite{Pol-tasi96}:
\be 
\label{brane-within-brane}
\mu_5 \int C {\rm Str} e^F
\Rightarrow \mu_5 \int d^6 x [C^{(2)} \wedge F \wedge  F]
\ee
which shows that non-zero values of $F_{67}, F_{89}$ can act
as a source term for $C^{(2)}_{05}$. The latter corresponds to
a D1-brane stretching in the 5 direction. In the above
equation 
\be C \equiv \oplus_n C^{(n)} \ee
and Str represents symmetrized trace. 

\tgap

The above observation leads us to look for non-trivial solutions of
six-dimensional SYM theory. 

We shall look for solutions which satisfy two conditions: 

(a) the $U(Q_5)$ gauge field should be 
independent of $x^{0,5}$, with
$A_0= A_5=0$. (This corresponds to the fact that
$x^{0,5}$ are Killing vectors in the supergravity solution.)  

(b) the solutions should preserve 1/2 of the supersymmetries, again
imitating the corresponding statement in supergravity.

In other words, we are looking for \underbar{static, stringy} 
solitons
of SYM$_6$ satisfying the BPS property.

Conditions (a) and (b) applied to SYM$_6$, leads to
\be
\delta_\epsilon \lambda \propto \Gamma_{ab} F^{ab} \epsilon =0
\ee
where $a,b$ run over $6,7,8,9$. It is easy to see that
this is equivalent to
\be
\label{instanton-equation}
F_{ab} = \epsilon_{abcd} F^{cd}, \; a,b,..= 6,..,9
\ee
where
\be
\label{instanton-susy}
\Gamma_{6789}\epsilon = \epsilon
\ee
These are nothing but instanton solutions of Euclidean
SYM$_4$. 

Note that equation \eq{instanton-susy} represents the only 
choice of unbroken SUSY consistent with condition (a). 

We now note the following points to bring out in more detail the
connection between these instantons and D1-brane
embedded in D5-branes.

(1) D1-branes break by half the sixteen unbroken supersymmetries
of D5-branes. To be precise, the supersymmetry breaking goes
as 
\be 
((2,2)+ (2',2'))_+ + ((2,2') + (2',2))_-
\to (2,2)_+ + (2,2')_-
\ee
where the representation labels correspond to the spacetime symmetry
group $SO(4)_{6789} \times SO(4)_{1234} \times SO(1,1)_{05}$. The
instanton solution, it can be checked, precisely breaks the same
supersymmetries as the D1-brane (recall \eq{instanton-susy}).

(2) The coupling in \eq{brane-within-brane}, together with the solution
\eq{instanton-equation}, imply that the source of the field $C^{(2)}$
is the Chern class of the four-dimensional gauge field
$F_{ab}$. Indeed the integral property of $F \wedge F$ exactly
corresponds to the quantization of D1-brane charge.  In other words,
one can easily see that the instanton action for a $Q_1$-instanton
solution is $Q_1/g_{YM}^2$. This agrees with the tension of $Q_1$
D1-branes, namely $Q_1/\gst$. (Recall that $\gst = g_{YM}^2$.)

(3) The ADHM construction \cite{ADHM-ref} of a $k$-instanton moduli
space for SU(N) Yang-Mills in $R^4$ is closely connected with
hypermultiplets in $SU(N) \times SU(k)$: namely, the ADHM equations
correspond to D-flatness conditions for the hypermultiplet fields
\cite{Dou95}. The solutions to the D-flatness condition of course
correspond to space of vacua, or moduli space, of the hypermultiplets.

\tgap
\hfill ---o--- \hfill
\tgap

\ni\underbar{Structure of moduli space}:

\ni We now discuss the structure of the $k$-instanton moduli space of
$U(Q_5)$ SYM theory on $T^4$. To begin with, note that $Q_1$ D1 branes
(along $x^5$) on $Q_5$ D5 branes ($x^{56789}$) can be T-dualized to
$Q_1$ D0 branes on $Q_5$ D4 branes ($x^{6789}$).  For $Q_5=1$, the
collective coordinates of the T-dualized system would correspond to
translation of $Q_1$ points on the $T^4$ along $x^{6789}$. Since the
D0-branes are {\sl identical}, (this corresponds to Weyl symmetry of
$U(Q_1)$) the $Q_1$ points should be unordered. The instanton
moduli space in this case should then be given by
\cite{Sen95-u,Sen95-note,Vaf95-instanton}
the space of $Q_1$ unordered points on $T^4$ is given by
\be
\label{unordered}
{\cal M} = \frac{(T^4)^{Q_1}}{S(Q_1)}
\ee
For $Q_5 > 1$, taking clue from the case $Q_5=2$
\cite{Vaf-Wit94}, one
arrives at the guess \cite{Vaf-Wit94,Vaf95-gas} 
\be
\label{moduli-space}
{\cal M} = [\frac{(T^4)^{Q_1Q_5}}{S(Q_1Q_5)}]
\ee
The notation $[\;]$ implies an appropriate
resolution of the orbifold and possible additional
factors corresponding to overall centre-of-mass motions.
For more details, see \cite{Dij98,Sei-Wit99,Mal-Moo-Str99,Dav-thesis}.

\tgap
\hfill ---o--- \hfill
\tgap

(4) One of the most quantitative evidences in favour of 
\eq{moduli-space}
is that the cohomology of the ${\cal M}$ agrees with
a U-dual \cite{Sen95-u} 
version of the cohomology of the D1-D5 system.
The argument goes as follows. 

A fundamental string with winding number $w_6$ and momentum
$p_6$ along the circle $x^6$ (say), can be mapped by using the
sequence of dualities $T_7ST_{6789}S$ to $w_6$ D2-branes along $x^{67}$
and $p_6$ D2-branes along $x^{89}$.  For our purpose here, we should
choose $w_6=Q_1, p_6=Q_5$ and make a further $T_{567}$ transformation
to get $Q_1$ D1-branes along $x^5$ and $Q_5$ D5-branes along
$x^{56789}$.

Now, BPS states of such a fundamental string simply correspond to
oscillator numbers $N_L =0, N_R = Q_1Q_5$. The number of
states $d(n)$ with $N_L=0, N_R=n$ is given by the standard
partition function formula (no $\bar q$ since $N_L=0$)
\be
\sum_n d(n) q^n = 256 \prod_n (\frac{1+q^n}{1-q^n})^8
\ee
Here 256 reflects the ground state degeneracy.
The dimension $d(Q_1Q_5)$, from the above formula
then gives the number of BPS states U-dual  to the ground
state of the D1-D5 system. 

If \eq{moduli-space} is true, then its BPS states should reproduce the
same number. The supersymmetric states of the moduli space
corresponds to its cohomology vector space $H^*({\cal M})$.
The dimension of this space indeed equals $d(Q_1Q_5)/256$
as required by U-duality (the factor 256 arises because
the hypermultiplet moduli space is based on gauge group
$U(Q_1) \times U(Q_5)$ rather than the $SU$ groups; it is
the latter that is dual to the instanton moduli space). 

More direct evidence has been provided for the
case of $Q_5=1$ in which case the five-dimensional
gauge theory is trivial and low energy degrees of freedom
indeed correspond to \eq{unordered}. 

\sgap

The ansatz about the moduli space leads to the prediction
that the low energy excitations are given by a conformal
field theory based on the manifold ${\cal M}$, as
the sigma-model flows in the infra-red to CFT. The  supersymmetry
${\cal N}=(4,4)$ of the sigma-model enhances in this limit
to ${\cal N}=(4,4)$ superconformal symmetry.

\sgap

\ni\underbar{Comment on resolution of orbifold, 
Fayet-Iliopoulos parameters and noncommutative geometry}

\sgap

The discussion in this section has not included for lack of space a
number of rather interesting issues related to resolutions of the
orbifold and marginal deformations of the SCFT. More thorough
discussions can be found in the original references mentioned earlier
or in \cite{Dav-thesis}. We have also not discussed the general issue
of stability of the bound state; it turns out that turning on the
marginal operators involved in the resolution of the orbifold
corresponds to turning on Fayet-Iliopoulos parameters in the D-term
equations \eq{dterm-1} and \eq{dterm-5}. The latter is known to remove
the singularity associated with the origin of the Higgs branch which
can be interpreted \cite{Sei-Wit99} as a D1-brane splitting off from
the bound system. In supergravity, on the other hand, a truly bound
state of D1 and D5 branes requires the introduction of a NS B-field
(without it, the mass of the bound state is the sum of the rest
masses, and a D1-brane can split off at no cost of energy). Using the
connection between NS B-field through noncommutative geometry (see
\cite{Sei-Wit99b} and references therein) between NS B-field and
Fayet-Iliopoulos terms, one can complete the picture (see
\cite{Dha-Man-Wad-Yog99} and references therein) of the stabilized
D1/D5 system in supergravity and gauge theory, including the moduli
space of deformations around them.

\sgap

{\bf Description of the SCFT degrees of freedom}

\tgap

The basic fields of the above-mentioned
SCFT  are $x^i_A(z,\bar z)$, $\psi^{a\alpha}_A(z)$ and 
$\bar \psi^{\dot a\dot \alpha}_A(\bar z)$. $A=1,2,\ldots Q_1Q_5$
denotes which-th copy of $T^4$. $i=6,7,8,9$ denote coordinate
labels on $T^4$. $a,\dot a$ denote spinor labels of
$SO(4)_I \equiv SO(4)_{6789}$ and $\alpha, \dot \alpha$ denote
spinor labels of $SO(4)_E$. The Lagrangian is given
by

\be
\label{cft}
S_0 = \int d^2 z[\half \del x^i_A \bar \del x^i_A +
\half \psi^{a\alpha} (z) \bar \del \psi_{a\alpha}
+ \half \psi^{\dot a\dot \alpha} (\bar z)  
\del \psi_{\dot a \dot \alpha}]
\ee

The orbifold in \eq{moduli-space} corresponds to
various twisted sectors corresponding to 
conjugacy classes of elements $g$
of the permutation group. It is well-known that
the distinct conjugacy classes of $S(m)$ are given
by the number distribution $n_i$ of cycles of various 
lengths $l_i$. These numbers satisfy

\be
\sum_i n_i l_i = m 
\ee

In our case $m = Q_1 Q_5$.

Thus, for $S(3)$ the distinct conjugacy classes can
be represented as $(1)(2)(3), (12)(3), (123)$.

The fields $x^i_A, \psi^{a\alpha}_A$ in a sector $(n_i, l_i)$
appear as a collection of $n_1$ strings of length $l_1$
plus  $n_2$ strings of length $l_2$, etc.
\cite{Dij-Moo-Ver-Ver96} For example,
in the maximally twisted sector, corresponding to
the maximal cycle of length $m$, the boundary 
condition on the bosonic coordinate is

\bea
\label{def-twist}
g: x^i_A(\sigma) &\to&  x^{(g),i}_A(\sigma) \equiv
x^i_{A+1}(\sigma) \\
&=&x^i_A(\sigma + 2 \pi)
\eea
where for $A=m$ we define $A+1=1$.

It is clear that the fields $x^i_A$ satisfying the
above boundary condition can be sewn together to form
a single field $\tilde x^i$ with periodicity $2\pi m$
(corresponds to $m$ times the length of the original
circle), defined by
\be
\tilde x^i(\sigma + 2\pi(A-1),t) \equiv x^i_A(\sigma,t),
\sigma\in [0,2\pi )
\ee
The sewn field will have a normal mode expansion:
\bea
\tilde x^i(\sigma,t) &=& (4\pi)^{-1/2} 
\sum_{n>0} [(
 \frac{a^i_n}{\sqrt n} e^{i n(-t+\sigma)/Q_1Q_5} 
\nn\\
&+& \frac{\tilde a^i_n}{\sqrt n} e^{i n(-t-\sigma)/Q_1Q_5} )
+ {\rm h.c.} ]\nn\\
\eea
The twist \eq{def-twist} acts on these oscillators as
\bea
\label{twist2}
g:a^i_n &\to& a^i_n e^{2\pi i n/Q_1Q_5} \nonumber\\
g:\tilde a^i_n &\to& \tilde a^i_n e^{-2\pi i n/Q_1Q_5}
\nonumber\\
\eea

\sgap

\ni\underbar{ Definition of the microstates}

\tgap

Let us now concentrate on the maximally twisted sector
alone.
The states $|i\rangle$ are now defined as
\cite{Dha-Man-Wad96,Dav-Man-Wad98} 
\be
\label{eq-state}
|i\rangle = \prod_{n=1}^{\infty}\prod_{i} C(n,i) (a^{i\dagger}_n
)^{N_{L,n}^i} (\tilde a^{i\dagger}_n )^{N_{R,n}^i} |0 \rangle 
\ee
where $C(n,i)$ are normalization constants ensuring unit norm of the
state. $|0 \rangle$ represents the NS ground state. The present
discussion is also valid in the Ramond sector, in which case the
ground state will have an additional spinor index but will not affect
the $S$-matrix. We have also suppressed for simplicity the fermion
creation operators which can be trivially incorporated.

\sgap

\ni{\bf Gauge invariance}

\tgap

The permutation group $S(m)$ arises as residual gauge symmetry
of the Yang-Mills theory, and the microstates
$|i\rangle$  above should
be invariant under it. Below we show how.

It is clear that the creation operators create KK
(Kaluza-Klein) momentum along $S^1$
(parametrized by $x^5$). The total left (right) moving KK momentum of
\eq{eq-state} (in units of $1/\tilde R, \tilde R \equiv Q_1Q_5 R_5$,
$R_5$ being the radius of the $S^1$) is $N_L$ ($N_R$),
where
\be
\label{total-n}
N_L = \sum_{n,i} n N^i_{L,n}, \quad N_R = \sum_{n,i} n N^i_{R,n} 
\ee
{}From \eq{twist2} and \eq{eq-state}, we see that
\be
g:|i\rangle \to \exp[\frac{2\pi i}{Q_1Q_5} (N_L-N_R) |i\rangle
\label{twist3}
\ee
Now, the total KK momentum carried by $|i\rangle$
is $p_5= (N_L - N_R)/(Q_1 Q_5 R_5)$. Quantization of the KK charge
requires that $p_5 = \hbox{integer}/R_5$, which implies that
\be
\label{twist4}
(N_L - N_R)/(Q_1 Q_5) = \hbox{integer}
\ee
Thus, using \eq{twist3} and the above equation, we find that 
the states $|i \rangle$ representing microstates of the
black hole are gauge invariant\cite{Dav-Man-Wad98}\footnote{
In \eq{total-n} we
have written down only the bosonic contribution to the 
KK momentum. If we wrote the total contribution of bosons
and fermions, then taking into account the
fact that an identical 
gauge transformation property holds for the fermionic oscillators
as in \eq{twist2}, we would arrive at the same 
conclusion as above, {\em viz}, that the state $|i\rangle$ 
is gauge invariant.}.

\sgap

\ni{\bf Entropy}

\sgap

It can be argued (see, e.g. 
\cite{Per}; also see below) that for entropy counting,
 the maximally twisted
sector dominates. Thus we are, roughly speaking, left with a free gas
of bosons and fermions moving in a large circle (of
$Q_1 Q_5$ times the original size).

Now we know that for a free gas of $N_B$ species of bosons and $N_F$
species of fermions, all moving to the left, in a one-dimensional box of
length $\tilde R$, the total energy is 
\be 
\label{1dgas}
E_L \equiv \frac{N_L}{R_5} \equiv 
\frac{\tilde N_L}{\tilde R} = \frac{\pi^2}{6}
\frac{\tilde R}{\beta_L^2}(N_B + \half N_F) = \pi^2
\frac{\tilde R}{\beta_L^2} 
\ee 
$\tilde R= Q_1 Q_5 R, \tilde N = Q_1 Q_5 N$. Using 
\be 
\del S_L/\del E_L = \beta_L(E_L) 
\ee 
we get 
\be 
\label{instentropy}
S_L= 2\pi\sqrt{E_L \tilde R}=2\pi\sqrt{N Q_1 Q_5} 
\ee 
reproducing the Bekenstein-Hawking result from free 1D gas!

Note that the entropy could also be computed by Cardy's
formula, as in \eq{cardy-formula}. The fact that
\eq{instentropy} gives the same result provides additional 
{\sl a posteriori} justification for considering
contribution only from the maximally twisted sector!

The above could appear to imply that the physics of the D1/D5 system,
at least at low enough energies, could be entirely described by a
``long string'' picture \cite{Mal-Sus96,Das-Mat96}.  We shall see
later that this expectation is not right: twisted sectors other than
maximal play a crucial role and the assumption of a long D-string
gives wrong coupling to bulk fields (Section 4). It is essential to go
back to the full orbifold SCFT and its deformations for a precise
understanding of the D1/D5 system.

\sgap

\ni\underbar{Density matrix interpretation}

\sgap

The above derivation of the entropy is
based on the assumption that  the
quantum black hole is represented by a density matrix
\be
\rho = \frac{1}{\Omega}\sum_i |i \rangle \langle i|
\ee
where the sum is over microstates which correspond to   
the given macroscopic charges $Q_1, Q_5, N$ 
that are reflected in the geometry. $\Omega=$
the total number of such microstates.

The density matrix reflects an averaging over microstates exactly as
we do in statistical mechanics. The deeper issue (on the lines of
ergodic theory) of how the density matrix may appear naturally in
classical time scales of observation (long compared to some typical
mixing times between states) merits detailed investigation.

We will see later that the density matrix ansatz correctly
reproduces Hawking radiation from the black hole.

\sgap

\ni{\bf Non-BPS entropy}

\tgap

For the non-extremal black hole \eq{non-extremal-5d} with no
``anti''-branes (i.e., $\alpha_{1,5}\to \infty, N_{\bar 1, \bar 5}=0,
\alpha_n < \infty$), we have both left and right moving gravitational
waves in the classical solution. This in the CFT should correspond to
exciting both $L_0$ and $\bar L_0$ 
(in the near-extremal case  $L_0
\equiv N \gg \bar L_0 \equiv \bar N $). If we assume that the left and
right moving oscillators do not interact, the total
degeneracy would be given by

\be
\Omega = \Omega_L (c, N) \times \Omega_R (c,\bar N)
\ee
so that
\be
\label{non-bps-entropy}
S = \log \Omega = 2 \pi \sqrt{Q_1Q_5}(\sqrt{N} + \sqrt{\bar N})
\ee
which is indeed the Bekenstein-Hawking entropy of the
non-extremal black hole as well! This is a surprise,
since there is no obvious non-renormalization
theorem for these systems which could protect the
density of states as the coupling constant is varied
from the D-brane regime to the supergravity regime
(see below the corresponding discussion for BPS states).

For later use, we note that the left- and right-movers
are separately represented by canonical ensembles
characterized by temperatures $\beta_L$ and $\beta_R$.
By \eq{1dgas} we see that

\be
\label{temperature}
\beta_{L,R}= \pi R_5 \sqrt{\frac{Q_1 Q_5}{N_{L,R}}}
\ee
We will find that the Hawking temperature is given by
\be
\label{hawk}
\beta_H = \half(\beta_L + \beta_R)
\ee

\sgap

\ni{\bf Weak and Strong couplings: BPS property}

\tgap

As has been indicated before, for the D1/D5 system
the supergravity description is reliable
for $g_{st} \to 0, Q_1, Q_5, N \to \infty$, such that
the scaled charges 
\be
\label{strong-thooft}
 \gst Q_1, \gst Q_5,  \gst^2 N \gg 1 
\ee
On the other hand, the D-brane description is valid when
\be
\label{weak-thooft}
 \gst Q_1, \gst Q_5,  \gst^2 N \gg 1 
\ee

Since the masses of BPS states do not change as we change the
coupling, the counting of states that we did using the
D-brane picture should continue to remain valid when
$\gst$ is increased from the range of
values \eq{weak-thooft} to \eq{strong-thooft}. It
is in this sense that we claim that we have a derivation
of the Bekenstein-Hawking formula.

What is surprising, however, is that we have agreement for non-BPS
entropy (as mentioned above) and, as we will see, for Hawking
radiation and absorption by near-extremal black holes. We will
analyze Hawking radiation and absorption now.

\section{\large Absorption/Emission}

\sgap

We saw in the last section that string theory provides 
an understanding of microstates of
the five dimensional black hole. In this section we will 
come back to some of the questions raised in the
introduction, and see if they can be addressed now that
we have a microscopic model of black holes.

We will first address the issue of absorption by a black hole. As we
remarked, classically the black hole only absorbs and does not
emit. We would like to see how this is interpreted in the microscopic
model. This would correspond to {\em an} explanation of a crucial
aspect of the event horizon.

Next we will turn on quantum mechanics and would like to
see how standard scattering processes described in terms
of the microscopic model gives rise to Hawking radiation.

Before we get to the microscopic understanding, let us
briefly review the (semi)classical calculations of
absorption/emission.

\tgap

\subsection{\large\bf Semiclassics of absorption/emission}

In this section we will show how to obtain black hole absorption
cross-section of various particles of type IIB supergravity.

We will mainly consider scalars. These can be minimal, namely if they
couple only to the five-dimensional Einstein metric and nothing
else. These are the simplest to discuss and we will consider them
first. There are other scalars which couple, in addition, to dilaton,
Ramond-Ramond potentials etc. and are non-minimal; we will consider
later a sub-class of these called fixed scalars.

\sgap

\ni{\bf Minimal scalar}

\sgap

The absorption cross-section and emission rate of a particular field
depend on how the field propagates and backscatters from the geometry
of the black hole. We will therefore look at the equation
of propagation of scalar fluctuations. 

Consider type IIB Lagrangian compactified on $T^5$
\cite{Mah-Sch93,Cal-Gub-Kle96}:
\bea
\label{IIBlag}
S_5 &=& \frac{1}{2\kappa_5^2} \int d^5 x \sqrt{-g} [ R - \frac{4}{3}
(\del_\mu \phi_5)^2   \nn\\
&~~& - \frac{1}{4} G^{ab} G^{cd}(\del_\mu G_{ac}\del^\mu G_{bd}
+ e^{2 \phi_5} \sqrt{G} \del_\mu B'_{ac}\del^\mu B'_{bd}) 
\nn \\
&~~& - \frac{e^{-4\phi_5/3}}{4} G_{ab}F^{(K),a}_{\mu\nu}
F^{(K),b \mu\nu} - \frac{e^{2\phi_5/3}}{4}
\sqrt{G}  G^{ab} H_{\mu\nu a}H_b^{\mu\nu}
-  \frac{e^{(4\phi_5/ 3)}}{12} \sqrt {G}  
H^2_{\mu\nu\lambda}]
\nn\\
\eea
Here $a,b,\ldots=5,\ldots,9, \, \mu,\nu,\ldots=0,1,\ldots,4$
\bea
\phi_5 &=& \phi_{10} - (1/4)\ln [{\rm det}_{ab} G_{ab}]  
\nn\\
H_{\mu\nu a} &=& (dA_a)_{\mu\nu} + \ldots,\, A_{a\mu}=
B'_{\mu a} + B'_{ab} A_{\mu}^{(K)b}
\nn\\
H_{\mu\nu \lambda} &=& (d\tilde B)_{\mu\nu\lambda} + \ldots,
\, \tilde B_{\mu\nu} =B'_{\mu\nu}+ 
 A_{[\mu}^{(K)a}A_{\nu] a} - A_\mu^{(K)a}B'_{a b}A_\nu^{(K)b}
\nn\\
F^{(K)a} &=& dA^{(K)a}
\eea
$A^{(K)a}$ are the KK vector fields. The terms denoted
by $\ldots$ represent ``shifts'' in the field strengths
which are not important for  either the background geometries
or fluctuations we are going to consider.

This Lagrangian can be obtained from type IIB Lagrangian in 10
dimensions simply by repeating compactification on a circle several
times \cite{Mah-Sch93} (can also be obtained by duality on 11
dimensional supergravity).

We will start with an example of a simple minimal scalar.

The near-extremal metric of the five-dimensional black hole
is a classical solution of the above Lagrangian. Consider
the 10-dimensional form of the metric. 
Let us consider an off-diagonal metric fluctuation on 
$T^4$  viz.

\be
ds^2|_{T^4} = f_1^{1/2}f_5^{-1/2} dx^a dx^b (\delta_{ab}
+ \kappa_5 h_{ab})
\ee
where $h_{ab}$ has only off-diagonal entries, say $h_{89}$.

With this rescaling, the field $h_{ab}$ is canonically normalized;
that is, the quadratic part of the Lagrangian involving $h_{ab}$
is

\be
S = \half \int d^5x \sqrt{-g} \del_\mu h_{ab} \del^\mu h_{ab}
\ee

It couples only to the five-dimensional metric, and is hence
a minimal scalar.

For the semiclassical analysis
\cite{Dha-Man-Wad96,Das-Mat96,Mal-Str96}, all we need is the equation for
propagation of the fluctuation $\varphi\equiv h_{ab}$ on the 
black hole metric, namely

\be
D_\mu \del^\mu \varphi =0
\ee

For the five-dimensional black hole metric discussed earlier
the equation becomes for the s-wave mode:

\be
[\frac{h}{r^3}\frac{d}{dr} (h r^3 \frac{d}{dr}) + f w^2]R_w(r) =0
\ee
where
\be
\varphi= R_w(r) \exp[-iwt]
\ee

The idea behind the absorption calculation is very simple. 
In terms of $\psi = r^{3/2} R$ the above equation becomes 

\be
[-\frac{d^2}{dr^2_*}  + V_w(r_*)]\psi = 0
\ee
where
\be
V_w (r_*) = -w^2 f + 
\frac{3}{4 r^2}(1 + 2 r_0^2/r^2 - 3 r_0^4/r^4)
\ee

The shape of the potential is given by (\fig{f-potential}).
\begin{figure}[!ht]
\begin{center}
\leavevmode
\epsfbox{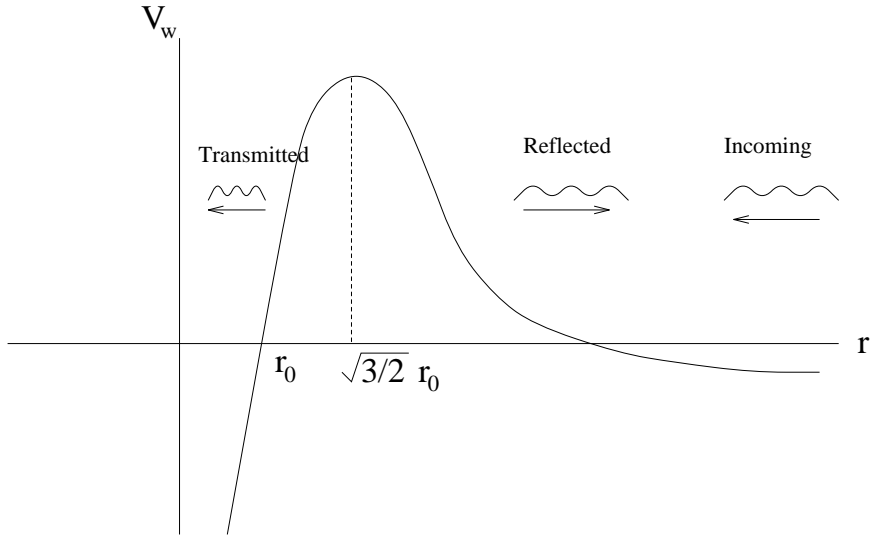}
\end{center}
\caption{Potential for minimal scalar}
\label{f-potential}
\end{figure}
Absorption is caused by the tunnelling of an incoming
wave into the ``pit of the potential''.

\tgap

\underbar{Near and Far solutions:}

\tgap

It is not possible to solve the wave equation exactly.  However, we
can devise near and far zones where the potential simplifies enough to
admit known solutions.  If the zones have an overlap region then
matching the near and far wave-functions and their radial derivatives
will provide the solution for our purpose. We will work
in the following range of frequency and parameters \cite{Mal-Str96}
\bea
r_0, r_n &\ll&  r_1,r_5
\nn\\
wr_5 &\ll& 1
\nn\\
w/T_{R,L}, r_1/r_5, r_0/r_n &\;& O(1)
\eea
$T_{R,L}=1/\beta_{R,L}$ have been defined in \eq{temperature}.
The far and near solutions will be matched at a point
$r_m$ such that
\be
r_0, r_n \ll r_m \ll r_1, r_5, \qquad wr_1 \ll r_m/r_1
\ee

\sgap

\underbar{Far zone ($r \ge r_m$):} 

\ni Here the potential $V_w$ becomes (in terms of 
$\rho=wr$)
\be
V_w(\rho)= -w^2(1 - \frac{3}{4 \rho^2})
\ee
This gives a Bessel equation, so that
\bea
\psi &=& \alpha F(\rho) + \beta G(\rho)
\nn\\
F(\rho) &=& \sqrt{\pi \rho/2}J_1(\rho),\quad 
G(\rho) = \sqrt{\pi \rho/2}N_1(\rho)
\eea
For $\rho\to \infty$ it is easy to read off the parts
proportional to $e^{\pm iw r}$.

\sgap
\underbar{Near zone ($r \le r_m$):}

\ni Here we have

\be
\frac{h}{r^3}\frac{d}{dr} (h r^3 \frac{d}{dr}R) 
+ [\frac{(wr_nr_1r_5)^2}{r^6}
+ \frac{w^2r_1^2 r_5^2}{r^4} ]R_w(r) =0
\ee
which is a Hypergeometric equation, with solution
\bea
R &=& A \tilde F + B \tilde G 
\nn\\
\tilde F &=& z^{-i(a+b)/2} F(-ia, -ib, 1-ia -ib,z)
\nn\\
\tilde G &=& z^{i(a+b)/2} F(-ia, -ib, 1-ia -ib,z)
\nn\\
z &=& (1- r_0^2/r^2)
\nn\\ 
a &=& w/(4 \pi T_R), b=w/(4 \pi T_L)
\nn\\
\eea
The temperatures $T_{R,L}$ are given by
\be
T_{L,R} = \frac{r_0}{2\pi r_1 r_5}e^{\pm \alpha_n}
\ee
These expressions agree with \eq{temperature}.

Now we impose the condition on the near solution that
the wave at the horizon should not have any outcoming
component: it should be purely ingoing
(no ``white hole''). This requires $B=0$.

\sgap

\ni\underbar{Matching}

We now match $R$ and $\frac{d}{dr} R$ between the near and
far regions at some point $r_m$ in the overlap
region.

This gives
\bea
\sqrt{\pi/2} w^{3/2} \alpha/2 &=& A e_1
\nn\\
e_1 &\equiv& \frac{\Gamma(1-ia-ib)}{\Gamma(1-ib)\Gamma(1-ia)}
\nn\\
\beta/\alpha &\ll & 1
\eea

\sgap

\ni\underbar{Fluxes}

The equation for $R$ implies
\be
\frac{d}{dr} {\cal F}=0
\ee
where
\be
{\cal F}(r) = \frac{1}{2i}[ R^* hr^3 dR/dr - {\rm c.c.}]
\ee

In order to find out what fraction of the flux gets
absorbed at the horizon, we compute the ratio
\be
R_1 = {\cal F}(r_0)/{\cal F}^{in}(\infty)= r_0^2 
\frac{a+b}{w|e_1|^2} w^3 \pi/2
\ee
where the superscript ``in'' indicates the flux calculated
from the ``ingoing'' part of the wave at infinity.

\tgap

\ni{\bf Absorption Cross-section:}

\tgap

In order to define absorption cross-section in the standard
way, one has to consider plane waves and not $s$-waves.
Using the fact that
\be
e^{-iwz} = (4\pi/w^3) e^{-iwr}Z_{000} + {\rm other\;partial\;waves}
\ee
and the standard definition of absorption cross-section
we get \cite{Mal-Str96}
\bea
\label{class-abs}
\sigma_{abs} &=& (4\pi/w^3) R_1 
\nn\\
\; &=& 2 \pi^2 r_1^2 r_5^2 \frac{\pi w}{2} 
\frac{\exp(w/T_H)-1}{(\exp(w/2T_R)-1) (\exp(w/2T_L)-1)}
\eea
In the $w\to 0$ limit, one gets \cite{Dha-Man-Wad96}
\be
\sigma_{abs} = A_h
\ee
where $A_h$ denotes the area of the event horizon.

\tgap

\ni{\bf Hawking radiation:}

\tgap

The semiclassical calculation of Hawking radiation
is performed through the standard route of
finding 
Bogoliubov coefficients representing mixing of
negative and positive frequency modes due to
evolution from  ``in'' to ``out'' vacua,
relevant for Minkowski observers existing
in the asymptotically flat regions at
$t=-\infty$ and $t=+\infty$ respectively
\cite{Haw75}.
The rate of radiation is given by
\be
\label{class-decay}
\Gamma_H = \sigma_{abs} (e^{w/T_H}-1)^{-1} 
\frac{d^4k}{(2 \pi)^4}
\ee
We will see below how $\Gamma_H$ and $\sigma_{abs}$
are reproduced in the D-brane picture.  

\tgap

\subsection{\large\bf D brane description}

\tgap

In this section we will discuss the case of minimal scalars.

We have seen that the entropy of the five-dimensional black hole can
be reproduced in the weak coupling regime by representing it as a
density matrix
\be
\rho= \frac{1}{\Omega}\sum_i | i \rangle \langle i |
\ee
where $| i \rangle$ represent the microstates. 

The picture of absorption in the microscopic
description \cite{Dha-Man-Wad96} is as follows. 
Consider throwing in a closed string mode, say
a minimal scalar $\varphi_n$, towards the D1-D5 configuration.
A microstate $|i\rangle$ 
will couple to such a fluctuation
in a certain way through some interaction
\be
S_{int} = \int d^2 z  \varphi_n|_B O_n(z, \bar z) 
\ee
and will get excited to a
different state $| f \rangle$ with the amplitude
\be
S_{if} = \langle f | H_{int} (| i \rangle| \psi_c\rangle)
\ee
where $|\psi_c \rangle$ represents the closed string mode. 
$\varphi_n|_B$ means here the value of the supergravity
mode on the brane.

Since we started from a density matrix description
for the initial state rather than individual microstates,
the probability of the process would be given by the
``unpolarized'' or ``inclusive''  expression:
\be
{\rm Prob}_{abs}= \frac{1}{\Omega}\sum_i \sum_f |S_{if}|^2
\ee
Note that the ``unpolarized'' transition probability 
corresponds to {\sl averaging} over initial states 
and {\sl summing} over final states. $\Omega$ is the
total number of initial microstates representing the
macroscopic charges of the black hole.

The most crucial ingredient in this calculation is to figure out what
$H_{int}$ (or $S_{int}$) is. In particular, given a particular
supergravity mode, what the corresponding operator $O(z,\bar z)$
is. This is the issue we discuss next.

\tgap

\underbar{Coupling supergravity modes to D-branes}\cite{Dav-Man-Wad98}

\tgap

We will work with the picture of microstates obtained
from the instanton moduli space described in the
last section.

As we remarked above, the coupling of a minimal scalar 
$h_{ij}$ to the
D-brane degrees of freedom is given in the form of the interaction 
\be
\label{one}
S_{\rm int}= \mu \int d^2 z h^{ij}|_B O_{ij}(z, \bar z) 
\ee
Here $h_{ij}|_B$ denotes the restriction of $h_{ij}$ to the
location of the SCFT, and $\mu$ is a number denoting the
strength of the coupling. (the indices $i,j$ in this section
will take values 6,7,8,9) 

The question is, what is the SCFT operator $O_{ij}(z, \bar z)$?

One way to determine it, of course, would be to
reanalyze the instanton moduli space or the hypermultiplet moduli
space with the metric of the $T^4$ deformed by $h_{ij}$. 
With the present level of technology this is not very feasible.

A simpler but more elegant approach is to appeal to {\sl
symmetries}. The steps involved are: \\
\ni (a) find the symmetries $\S$ of the bulk,\\ 
(b) find how (all, or a part of) these
symmetries appear in D-brane world-volume and consequently how they
act on the variables of the SCFT,\\ 
(c) find how $h_{ij}$ transforms
under $\S$, and \\
(d) demand that $\o_{ij}$ should transform under the
{\sl same} representation of the symmetry group $\S$ when it acts on
the SCFT. 

The last step arises from the fact that $h_{ij}(\z)$ in
\eq{one} is a source for $\o_{ij}$.

\sgap

We have already indicated that the symmetries $\S= SO(4)^I\times
SO(4)^E$ of the bulk theory (tangent group of the 4-torus and rotation
in the transverse space respectively) appear naturally in the SCFT of
the D-brane world volume as well. The $SO(4)^I$ part is obvious;
$SO(4)^E$ appears as the R-parity group.

Let us now apply steps (c) and (d) above in the context of this
symmetry group $\S$.

The field $h_{ij}$ (symmetric, traceless) transforms as $(\bf 3, \bf
3)$ under $SO(4)^I\equiv SU(2)^I \times SU(2)^I$ and as $(\bf 1, \bf
1)$ under $SO(4)^E \equiv SU(2)^E \times SU(2)^E$. 

Now there are at least three possible SCFT operators which belong to
the above representation of $\S$:
\bea
\o_{ij} &=& \del x^i_A \bar \del x^j_A 
\nonumber \\
\o'_{ij} &=& \psi^{\alpha}_{aA}(z) \sigma_i^{a \dot a}
\bar\psi^{\dot \alpha}_{\dot a B}(\bar z) 
\psi_{\alpha,bA}(z) \sigma_j^{b \dot b}
\bar\psi_{\dot \alpha,\dot b B}(\bar z)  
\nonumber \\
\o''_{ij} &=& \psi^{\alpha}_{aA}(z) \sigma_i^{a \dot a}
\bar\psi^{\dot \alpha}_{\dot a A}(\bar z) 
\psi_{\alpha,bB}(z) \sigma_j^{b \dot b}
\bar\psi_{\dot \alpha,\dot b B}(\bar z)  
\label{two}
\eea
The spinor labels are raised/lowered above using the
$\epsilon^{\alpha\beta},\epsilon_{\alpha\beta}$ symbol. The
$\sigma_i$'s denote the matrices : $(1, i\vec \tau)$.  The last two
operators differ only in the way the $S(Q_1Q_5)$ labels are
contracted. All the three operators should be symmetrized
$(i,j)$ and projected onto the traceless part.

The complete list of operators with the same transformation property
under $\S$ contains, in addition, those obtained by multiplying any
of the above by singlets. These would necessarily be irrelevant
operators, but cannot be ruled out purely by the above symmetries.

It might seem `obvious' that the operator $\o_{ij}$ should be the
right one to couple to the bulk field $h_{ij}$. However, the simplest
guesses can sometimes lead to wrong answers, as we will see
later for fixed scalars, where it will turn out that the operator
$\del x^i_A \bar \del x^i_A$ is far from being the right one to couple
to $h_{ii}$ (trace). We proceed, therefore, to look more closely
into symmetries to lift the degeneracy of the operators.

\sgap

\ni{\bf Fixing the operator using near-horizon Symmetry}

\sgap

It has been conjectured recently that if one takes the large $gQ$
($g=\gst, Q=Q_1,Q_5$) limit (which corresponds to the near-horizon limit in
supergravity and the large 'tHooft coupling limit in the gauge-theory
of the brane world-volume), a powerful correspondence
\cite{Mal97,Wit98-ads,Gub-Kle-Pol98} can be built between the physics
of the bulk and the physics of the branes at the boundary
(for a review, see 
\cite{Aharony}).
For the scaling of various quantities in this limit, see
the references just mentioned.

Let us consider for a moment the supergravity solution
for the D1/D5 system \eq{6d-black-string}. 
In the limit $gQ_{1,5} \gg 1$, we have
$r_{1,5} \gg l_s$. It is easy to see that the metric becomes
\be
\label{near-horizon}
ds_{10}^2 = ds_{AdS_3}^2 + ds_{S_3}^2 + ds_{T^4}^2 
\ee
where
\bea
ds_{AdS_3}^2 &=& \frac{r^2}{R^2}(-dudv) + \frac{R^2}{r^2}dr^2 
\nn\\
ds_{S_3}^2 &=& R^2 d\Omega_3^2
\nn\\
ds_{T^4}^2 &=& \frac{r_1}{r_5}(dx_6^2 + dx_7^2 + dx_8^2 + dx_9^2)
\eea
In the above, $R^2 \equiv r_1 r_5$.

In order to arrive at the black hole solution
\eq{extremal-5d},\eq{non-extremal-5d}, we have to first
compactify $x^5$. In terms of the near-horizon
limit, this turns
\cite{Mal-Str98} AdS$_3$ into a BTZ black hole
(with zero mass and angular momentum; turning
on momentum modes along $x^5$ corresponds to mass
and angular momentum of the BTZ solution).
In the SCFT, this corresponds to
switching from the NS sector to the Ramond sector. For our purposes
here it will be enough to continue to work with the AdS$_3$
description since the local operators of interest here can be shown to
have the same symmetry properties irrespective of whether they belong
to the Ramond or the NS sector. The ground state of the Ramond sector
is degenerate, as against that of the NS sector; this degeneracy would
be reflected in our construction of the black hole state --- however,
this would not affect the $S$-matrix relevant for absorption and
emission.  

It is easy to see that the symmetry group  
of the solution \eq{near-horizon} is enhanced to 
\be
\label{enhanced}
\S \to \S'=SO(4)^I
\times SO(4)^E \times SU(1,1|2) \times SU(1,1|2)
\ee 

The factor $SO(4)^I$ appears as
before. The group $SO(4)^E$ corresponds now to the isometry group of
$S^3$. The bosonic part of $SU(1,1|2)\times SU(1,1|2)$ arises as the
isometry group of $AdS_3$ (which is the $SL(2,R)$ group manifold).
The $SU(2)$ part which transform the fermions among themselves, and
the ``off-diagonal'' part of $SU(1,1|2)$, are a consequence of
${\cal N} = (4,4)$ supersymmetry of this compactification.

On the SCFT side, the $SO(4)_{I,E}$ groups have actions as explained
before.  The $SU(1,1|2)$ is identified with the subgroup of the
superconformal algebra generated by $L_{\pm 1,0}, G^{a\alpha}_{\pm
1/2}$ (the other $SU(1,1|2)$ involves $\bar L, \bar G$;.  
indices have been explained above \eq{cft}).
 
\sgap

Let us now apply steps (c) and (d) of Method 2 to this enhanced
symmetry group $\S$. How does $h_{ij}$ transform under $SU(1,1|2)$?

\sgap
\vfill\eject

\ni\underbar{Short multiplets of $SU(1,1|2)$}

\sgap

\begin{tabular}{|| l |  l |  l |  l ||}
\hline
States                  & $j$   & $L_0$         & degeneracy    \\
\hline
                        &       &               &               \\
$|\rangle$              & $h$   & $h$           & $2h + 1$      \\
                        &       &               &               \\
\hline
                        &       &               &               \\
$\bar G^1_{-\half}|\rangle,
G^2_{-\half}|\rangle$   &$h-
                          \half$& $h + \half$   & $2h + 2h$     \\
                        &       &               &               \\
\hline
                        &       &               &               \\
$\bar G^1_{-\half}
G^2_{-\half}|\rangle$   &$h-1$  & $h +1$        & $2h -1$       \\
                        &       &               &               \\
\hline
\end{tabular}
 
\sgap

Here $j$ denotes the $j$-value of R-parity group $SU(2)$
of $SU(1,1|2)$. Since our minimal scalar $h_{ij}$ is a singlet of
$SO(4)_E$, it has $j=0$. The fact that it is
a massless supergravity mode, leads to $L_0=1$.
{}From the table, such a multiplet corresponds to $h=\half$
and the field $h_{ij}$ fits into the middle row.
For $h=\half$, the middle row corresponds to the ``top''
component of the supermultiplet, since it is annihilated
by the raising operators $\bar G^1_{-\half},G^2_{-\half}$. 

{}From our list of candidate operators $\o_{ij}, \o'_{ij}$ and
$\o''_{ij}$:\nl
{\sl only $\o_{ij} = \del x^i_A \bar \del x^j_A$ belongs to
this representation.}

For a more detailed matching of various supergravity moduli
with short multiplets of the SCFT, see 
\cite{deB98,Dav-Man-Wad98,Lar-Mar99,Dav-Man-Wad99}.

\sgap

\ni{\bf S-matrix}

We have found above that
\be
\label{interaction} 
S_{\rm int} = \frac{\mu}{2}  \int d^2 z\; 
\left[ h_{ij} \del_z x^i_A \del_{\bar z} x^j_A \right]
\ee
We have omitted a factor of effective string
tension appearing in front of both $S_{\rm int}$ and $S_0$
(in \eq{cft}), since the
factor cancels in the $S$-matrix between the interaction Lagrangian
and the external leg factors.  However, the value of $\mu$ {\sl is
important} to determine since the absorption cross-section and Hawking
radiation rates calculated from the SCFT depend on it.
We do not have space to detail the argument but
the quantitative version of Maldacena conjecture
demands  that $\mu=1$. 

Let us now restrict our attention to the maximally
twisted sector of the orbifold Hilbert space, as 
in the case of the entropy calculation. This would imply,
as before, that the fields $x,\psi$ live on a large
circle, of radius 
\be
\tilde R= Q_1 Q_5 R_5
\ee

Using the interaction Hamiltonian obtained above,
and considering the example of $h_{89}$, we get
for the process 
\be
h_{89}(w,0) \to x^8_L(w/2,-w/2) + x^9_R((w/2,w/2)
\ee
(the numbers in parenthesis denote $(k_0, k_5)$)

\be
\label{s-matrix}
S_{if} = \frac{\sqrt{2} \kappa_5 w_1 w_2 \tilde R 
\delta_{n_1,n_2} 2\pi
\delta(w-w_1 - w_2)}{\sqrt{w_1 \tilde R
w_2 \tilde R w V_4}} \sqrt{N_{L,n_1}^8} 
\sqrt{\tilde N_{R,n_2}^9}
\ee
$V_4$ = volume of the noncompact space (box normalization).
The notation $N^i_{L,n}$ and $N^i_{R,n}$ denotes number
distribution of oscillators with left- and right-moving
momentum $n$ respectively (see \eq{eq-state}). The
factors of $\sqrt{N}$ appear from the identity
\be
\langle N-1 | a |N \rangle = \sqrt{N}
\ee
where $|N\rangle = (a^\dagger)^N/\sqrt{N!} |0\rangle,
[a, a^\dagger]=1$.

\tgap
\hfill ---o--- \hfill
\tgap

\ni{\bf Seeing Hair}: Eqn. \eq{s-matrix} shows that the S-matrix
describing emission (or absorption) of waves of a given frequency $w$
contain information about the number distribution of quanta of
frequencies $w/2$.  By repeating this for all $w$ (subject to the
overall condition that the frequencies are not too high), we can get
the (limited) information $\sqrt{N_L(w/2) N_R(w/2)}$ for all such $w$.
If, however, we are interested in the ``inclusive'' processes (like in
the case of unpolarized cross-sections for standard particle physics
experiments), then one computes probabilities by taking a
modulus-square of the S-matrix, {\em sum} over the final states and
and {\em average} over the initial states.

\tgap
\hfill ---o--- \hfill
\tgap

Thus, the probability of absorption of the quantum of frequency $w$ is
\bea
{\rm Prob}_{\rm abs}
&=& \Omega^{-1}\sum_{i,f}| S_{if} |^2
\nn\\ 
&=&  \frac{\tilde R T}{V_4} \kappa_5^2 w 
\langle N_L(w/2) \rangle\; 
\langle  N_R(w/2) \rangle
\nn\\
\eea
Here $T$ is the length of the time-direction.

The decay probability is obtained by considering
the reverse process. One gets
\bea
{\rm Prob}_{\rm decay}
&=& \frac{1}{\Omega'} \sum_{i,f}| S_{if} |^2
\nn\\ 
&=&  \frac{\tilde R T}{V_4} \kappa_5^2 w 
\langle N'_L(w/2) \rangle\; 
\langle N'_R(w/2) \rangle
\eea
where the primed number distribution in the last line 
refers to the final state. $\Omega'=$ total
number of final microstates. 

For the mode $w/2 = n_1/\tilde R$, $N_{L,R}(n_1)
= N'_{L,R}(n_1)+1$. The classical absorption probability
should be compared with Prob$_{\rm abs} - $ Prob$_{\rm
decay}$ of the string calculation.

$\sigma_{abs}$ is defined by 
\bea
\sigma_{abs} \times {\rm speed\, of\, particles} &\times& 
{\rm particles/volume} \equiv {\rm Rate}
\nn\\
&\equiv& {\rm Prob_{abs}}/T
\nn\\
\eea

Here, particles have speed = 1.  Also number of particles particles
per unit volume = $1 /V_4$ (box normalization).

Putting in all the above, we get
\be
\label{abs}
\sigma_{abs} = 2 \pi^2 r_1^2 r_5^2 \frac{\pi w}{2} 
\frac{\exp(w/T_H)-1}{(\exp(w/2T_R)-1) (\exp(w/2T_L)-1)}
\ee
which is the same expression as we had obtained
semiclassically.

The decay rate is given by
\be
\Gamma = {\rm Prob}_{\rm decay}\frac{V_4}{{\tilde R T}} 
\frac{d^4k}{(2 \pi)^4}
\ee
giving
\be
\label{decay}
\Gamma_H = \sigma_{abs} (e^{w/T_H}-1)^{-1} 
\frac{d^4k}{(2 \pi)^4}
\ee
which exactly reproduces the semiclassical result.

\sgap

We summarize this section by making the following comments:

\tgap

1. {\bf The black hole is a black body:} This {\em nearly} completes
the derivation that so far as the Hawking radiation of minimal scalars
is concerned, the emission from the black hole under discussion is
really that from a black body. The reason we say ``nearly'' is that
the strength of the interaction Hamiltonian is still not
determined from first principles, but rather by using
the postulate of the AdS/CFT correspondence \cite{Dav-Man-Wad98}.

\tgap

2. {\bf The black hole arrow of time is the same as the thermodynamic
  arrow of time}: In the limit $\hbar\to 0$, the enhancement factor
$\Omega'/\Omega$, representing the ratio of absorption versus emission
probabilities, blows up and the decay rate goes to zero in precisely
such a way that the absorption cross-section is finite and reproduces
the classical calculation precisely. This is how a black hole
classically only absorbs and does not emit. As we see, the explanation
comes merely from the way we calculate the inclusive processes
here, by summing over the final states and averaging over the
initial states; this is the assumption of randomization or
ergodization at the quantum mechanical level, giving rise to
thermodynamics.

\tgap

3. {\bf Randomization}: As emphasized above, we have represented the
state of a black hole by a density matrix in stead of in terms of any
specific microstate. The assumption made here is that of randomization
standard in any thermodynamic system.  In most simple thermodynamic
systems, we have a mechanism in mind how some interaction,
characterized by some time scale, between the microstates causes a
hopping between them, leading ultimately to density matrices. In the
case of the black hole microstates, as long as we are in the conformal
field theory description, the microstates described above are strict
eigenstates and there is no mixing between them. In order to see the
mixing, we need to go away from the infra-red fixed point and go back
to the effective sigma-model describing the gauge theory of the D1/D5
system. It will be extremely interesting to estimate the
time scale of the mixing from this and to understand how
a semiclassical calculation of Hawking radiation bypasses all
this and anticipates randomization in an in-built way.

\tgap

\subsection{\large\bf Fixed scalar}

\tgap

The above example describes the emission/absorption
of any scalar that couples only to the five-dimensional
Einstein metric. It is clear from the type II
Lagrangian in five dimensions
\eq{IIBlag} that not all scalars
are that way. For example, consider scalar
fluctuations described by
\be
(ds^2_{10})_{T^5}= e^{2 \nu_5} dx_5^2 +
e^{2 \nu}(dx_6^2 + dx_7^2 + dx_8^2 + dx_9^2)  
\ee
The field $\nu$ appearing above, and 
\be
\lambda = \frac{3}{4}\nu_5 - \half \phi_5
\ee
are examples of ``fixed'' scalars, which couple to the
KK vector field strengths and to the RR $B'$-field.

The wave equations are fairly complicated. The
solutions are obtained 
\cite{Kra-Kle96,Kra-Kle97,Marika} by matching behaviours
of solutions that are valid in a near region, 
intermediate region and a far region. The low
energy absorption cross-section that follows
vanishes as $w\to 0$.   

\sgap

\underbar{D-brane picture}

The first attempt at calculating the absorption/emission
in the D-brane picture was made
\cite{Cal-Gub-Kle96} by guessing the
following form of $S_{\rm int}$:
\be
S_{\rm int}= \int d^2 z[ \lambda|_B(\o_{3,1} +
\o_{1,3} + \o_{2,2})+ \nu|_B(\o'_{3,1} +
\o'_{1,3} + \o'_{2,2})]
\ee
The subscripts refer to values of $h,\bar h$.
This form of the interaction was guessed by 
imagining the degrees of freedom of the D1-D5
system to be those of a D-string and coupling it
to supergravity through a Dirac-Born-Infeld action.

The absorption/emission rates obtained thus were
at variance with the semiclassical calculation.

By applying the method we described above in the
context of the minimal scalars, namely by using
the near-horizon symmetry, we find that only
the (2,2) operators are allowed. Since the earlier
discrepancy was caused by the coupling to (1,3)
and (3,1) operators, we get agreement between 
semiclassical calculation and D-brane picture
\cite{Dav-Man-Wad98}. 

\tgap

\section{\large Discussion}

\tgap

We have touched on several open problems in the course of this review.
We have not had time to go into some others which I find
rather exciting. I will end this discourse with a short description of
some of these.

\tgap

\subsection{\large\bf Correspondence principle}

In our previous discussion we described our understanding of the
physics of the D1/D5 system and the five-dimensional black hole in
terms of D-brane microstates.  It is clear that supersymmetry has
played an important role in the entire discussion.  The question that
naturally arises is: how essential is the role of supersymmetry?
Another related point is that the understanding is rather
too detailed for comfort and any universality, if at all, is
fairly non-obvious.

An alternative way to address the question is to ask how generic is
the fact that black holes can be understood as states in a string
theory. Does a very massive string state always give rise to a black
hole irrespective of whether or not supersymmetry is present? 
To answer this, let us consider an elementary string state of
mass $M$ and increase the coupling from the string regime ( $r_h
\ll l_s$) to the supergravity regime $r_h \gg l_s$
(by $r_h$ we mean the radius of the horizon). It was emphasized
by Susskind  \cite{Susskind,Halyo}
that for the 4D Schwarzschild black hole, the entropy
formula (as a function of mass) given by the string theoretic
expression  
\be
\label{string-entropy} 
S_{\rm string}
\sim \sqrt{N} \sim \sqrt{\alpha'} M 
\ee
has a different functional form from the Bekenstein formula 
\be
\label{bekenstein-entropy}
S_{\rm BH} \sim G_N M^2,
\ee
A straightforward
identification of the black hole as a state in a string theory
therefore does not seem to be feasible. Susskind
argued, however, that the string theoretic entropy is 
calculated at string perturbation theory (in fact
at the tree level) and such 
a calculation need not be valid in the supergravity regime.
It should not come as a surprise, therefore, if the density of states
undergoes a change in form through renormalization.

Horowitz and Polchinski \cite{Hor-Pol} argued that a crucial
test of this idea (whether an elementary string collapses into a black
hole or not) is to find a region of overlap between the regions of
validity of the string description and the (super)gravity description
and to see if the formulae agree there. They found (upto a constant of
order one) a correspondence point $g=g_c$ below which the string
description should be valid and above which the gravity description
should be valid. They found in a wide variety of cases that the two
expressions for entropy match at the correspondence point, except
possibly upto a numerical constant.  This establishes a correspondence
principle for the scenario of a string collapsing into a black hole.

We quote   below the simplest case considered in \cite{Hor-Pol}, that
of the 4-D Schwarzschild black hole.  The functional forms of the
entropy, in the string and in the gravity regime, are different, as
mentioned above. Let us find the respective regions of validity of the
string and the gravity pictures. Clearly, for the gravity picture to
be reliable, the Schwarzschild radius $r_h$ should be large compared
to the string length $l_s = \sqrt{\alpha'}$, that is
\be
G_N M \sim l_s^2 \gst^2 M \gg l_s 
\Rightarrow M \gg M_c \equiv \frac{1}{\gst^2 l_s}
\ee
The string description, on the other hand, should be
valid at weak coupling $\gst$ where the mass of the string
will be smaller than $M_c$.  The transition point is given
by
\be
M= M_c = \frac{1}{\gst^2 l_s}
\Rightarrow  \gst = (l_s M)^{-1/2}
\ee
If we evaluate the string entropy and the Bekenstein-Hawking
entropy at this point we get
\be
S_{\rm string} \sim l_s M_c \sim  1/\gst^2
\ee
and 
\be
S_{\rm BH} \sim l_s^2 \gst^2 M_c^2 \sim  1/\gst^2
\ee
Thus the two distinct entropy formulae match at
the transition point upto a possible numerical
constant. The numerical constant cannot be fixed
more accurately since the correspondence point itself
is known only upto a numerical constant.

Having seen that the entropies \eq{string-entropy} and
\eq{bekenstein-entropy} go over to each other at the transition point,
one naturally wonders about the {\em mechanism} of the transition
between the two behaviours. In particular, whether it is possible to
understand \eq{bekenstein-entropy} by including the effect of
interactions in the string picture.  Such an attempt has been made in
\cite{Hor-Pol-a,Dam-Ven}. The idea in these references is that it is
the gravitational self-interaction of the strings that renormalizes
the density of states to convert it into the one appropriate for the
black hole.

It is very important to understand in detail the physics of the above
transition since it may provide some universal clues to gravitational
collapse into a generic black hole. An interesting
direction of investigation would be to try to understand such
a collapse  in the context of D-brane black holes
(extremal as well as non-extremal)\footnote{%
It should be emphasized that even in the BPS cases where the
density of states is not renormalized (or in some
non-BPS cases where, as mentioned in the next subsection,
the renormalization is only by a numerical factor) there
are non-trivial features of collapse into a black hole which bear
investigation. For example, a string state at weak
coupling is unlikely to satisfy no-hair theorems
(see, e.g., \cite{Man-Wad95,Dav-Dha-Man-Wad96}) whereas
in the domain of validity of classical supergravity,
black holes are expected to satisfy no-hair theorems
(see \cite{Lar-Wil}, though, for possible counterexamples).
Such change of behaviour, described possibly by an
order parameter related to some ``hair'', may be gradual
or in the form of a phase transition. We use below the
phrase ``string-black hole transition'' keeping  such
possibilities in mind.}. Two of these black holes have been
discussed extensively in the literature: the five-dimensional black
hole described in this review and the seven dimensional
black hole  whose BPS limit is the (wrapped) D3-brane. These systems
(more precisely their near-horizon counterparts, related to
$AdS_3\times S^3 \times T^4$ and $AdS_5 \times S_5$ respectively) have
a dual description: in terms of (a) string theory/supergravity and (b)
gauge theory/conformal field theory. The discussion of a string-black
hole transition would seem to appear naturally in the
string/supergravity description, but should have interesting
correspondence with the phase structure of the gauge theory. In the
context of D3-branes, the confinement/deconfinement transition of the
world-volume gauge theory has been related \cite{Wit98-thermal} to
a Hawking-Page \cite{Haw-Pag} phase 
transition in $AdS_5$ gravity. It would be
interesting to compare this transition with a possible string-black
hole transition in this system. The Hawking-Page transition occurs as
one varies the temperature whereas the string-black hole transition
occurs as one varies $\gst$. Both cases involve, in a sense, collapse of
strings into black holes: in the thermal case, strings in empty
AdS space condense to form a black hole whereas in the other case
strings in flat space condense to form a black hole.  
Thermal phase transitions have also been described
\cite{Mal-Str98,Dav-Man-Vai-Wad99} in the
AdS$_3$/CFT$_2$ framework which arises in the context of the
D1/D5 system. The possibility of a connection with
a string-black hole transition would be interesting
to explore in this case as well.
 
Another possible insight into the nature of the string-black hole
transition might be obtained from a study of stable non-BPS states
\cite{Sen-stable}.  Typically, generic massive string states that are
non-BPS have a decay width which increases as $\gst$ increases. These
states therefore get mixed up with string states at other mass levels,
thus complicating the issue of how to ``identify'' these states as
$\gst$ goes up. For the stable non-BPS states, this problem is not
there, and in principle they can be identified at strong coupling as
well. In a number of cases they have been identified and their masses
at strong coupling have been obtained exactly using duality arguments
\cite{Sen-stable}. These therefore provide examples of exact mass
renormalizations and it may be instructive to compare with the mass
renormalizations envisaged in the string-black hole transition.

\tgap

\subsection{\large\bf Obedient non-supersymmetric black holes}

A rather important aspect of nonsupersymmetric black holes is the
existence of examples where string/D-brane quantities agree with
results from (super)gravity without the help of any obvious
supersymmetry non-renormalization theorems. These are important to
study since they may provide us with a principle other than
supersymmetry to understand these black holes.  We mention briefly a
few examples (these are not exhaustive):

\begin{itemize}
  
\item{(1)} We have already mentioned in the context of the
  non-extremal black hole in five dimensions that the D-brane entropy
  formula \eq{non-bps-entropy} agrees with the semiclassical formula
  although the system is non-BPS.
  
\item{(2)} The agreement for the absorption cross-section and Hawking
  rate between the D-brane calculation and supergravity is also
  surprising since they refer to near-extremal black holes. It is not
  entirely clear for what range of energies above extremality (and for
  what precise reason) it should be possible to extend BPS
  non-renormalization arguments.
  
\item{(3)} There are examples of elementary heterotic string states
  \cite{Dab-Man-Ram} which are non-BPS (although extremal), but whose
  mass renormalizations are bound, to all loop order, by $1/M$ (rather
  than by $M$ as one would expect normally). This explains why the
  string entropy of these states matches, in the sense of
  \cite{Sen-entropy}, the semiclassical entropy of the stretched
  horizon of the corresponding black hole solution \cite{Gar-Hor-Str}.
  The mechanism for the existence of the bound, however, is rather
  technical and very stringy, and merits a simpler understanding.
  
\item{(4)} There is a type I black hole \cite{Dab}, closely related to
  the 5D black hole discussed in these lectures, but with no
  supersymmetry, whose entropy agrees with the Bekenstein-Hawking
  formula.  A   large $N$ (where $N$ refers to $Q_1,Q_5$) argument
  towards understanding this result has been presented in \cite{Barbon}.
  
\item{(5)} For a D0-D6 black hole  the ADM mass formula is precisely
  reproduced \cite{Dha-Man} in the gauge theory description of the
  system. This is surprising since the black hole is nonsupersymmetric
  and the two mass formulae are calculated at two different regimes of
  coupling.
  
\item{(6)} In the context of scattering involving a system of D0- and
  D2-branes \cite{Lif-Mat} or D0- and D6-branes
  \cite{Kes-Kra,Dha-Man}, one finds agreement between a matrix theory
  calculation and supergravity calculation in the limit when the D0-
  charge is much larger than the D2- (respectively D6-) charge
  (equivalent to a large boost along $x^{10}$ of M-theory).
  Ordinarily, one could say that this is an example of effective
  restoration of supersymmetry by a large boost (this is similar to
  the fact that in \eq{approach-extremal} extremality is equivalent to
  a large boost $\delta$). However, the agreement in question here is
  in the static potential between branes which does not even exist for
  BPS branes. The agreement here, therefore, does not automatically
  follow from supersymmetry.

\item{(7)} The entropy of the near-extremal D3-brane, as calcaluted
  from the world-volume gauge theory, matches \cite{Kle-Pee}, upto a
  factor $4/3$, the entropy calculated from supergravity. This is
  related to the demonstration in the context of AdS/CFT
  correspondence, that the entropy of an AdS-Schwarzschild black hole
  in five dimensions is exactly reproduced from SYM theory at finite
  temperature, upto a constant of proportionality
  \cite{Wit98-thermal}. Although the agreement is not exact, it is
  quite remarkable that, unlike in the Schwarzschild case mentioned in
  the last section, the functional form of the entropy is the same at
  both ends.
  
\item{(8)} The leading term in the high temperature partition function
  of a BTZ black hole agrees \cite{Dav-Man-Vai-Wad99} with the partition
  function of the two-dimensional CFT of the brane world-volume. The
  interesting point is that the agreement includes the case of zero
  angular momentum which is a completely nonsupersymmetric
  configuration. The agreement is exact and there are no factors.

\end{itemize}

It is clear that there would be considerable progress towards
understanding nonsupersymmetric black holes and nonsupersymmetric
gauge theory if one can tie up the above (and similar other) examples
alongwith the observations in the last section on the
phase transition of a string state into a black hole.

\tgap

\subsection{\large\bf The AdS/CFT conjecture and Complementarity}

\tgap

One of the important aspects of black hole physics is the issue of
complementarity, which says, very roughly speaking, that the physics
outside the horizon can encode information of ``stuff'' inside that
``makes'' the black hole.  
\begin{figure}[!ht]
\begin{center}
\leavevmode
\epsfbox{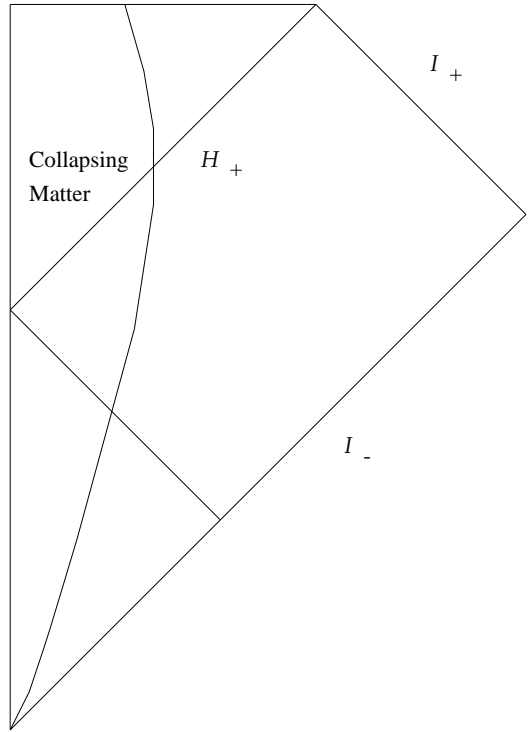}
\end{center}
\caption{Collapsing black hole}
\label{f-collapse}
\end{figure}
In a way the AdS/CFT correspondence
\cite{Mal97,Wit98-ads,Gub-Kle-Pol98} captures some
of the essence of complementarity. To elaborate, the Hilbert
space of the gauge theory/CFT on the brane-world-volume
contains information about the microstates of the black hole.
Normally one does not suppose supergravity states 
outside the horizon to carry information about
these microstates. However, supergravity states in the
near-horizon AdS geometry, by virtue of the AdS/CFT correspondence, 
are in one-to-one correspondence with the spectrum of operators
of the gauge theory/CFT. This implies that physics outside
the horizon can, after all, encode complete information about
gauge theory fluctuations which may in a sense be
regarded as degrees of freedom inside the black hole.
It is important to mention, however, that the holographic correspondence
has some crucial nonlocal features in it. For example,
local (delta-function) boundary fluctuations on the brane,
propagated to the bulk by means of the boundary-bulk Green's 
function \cite{Wit98-ads}, have nonlocal support. Similarly, local degrees
of freedom in the bulk get holographically projected to
nonlocal fluctuations on the boundary. This is also
reflected in the nonlocality of the operator algebra \cite{Banks}
on the boundary. Such nonlocality accords with the
expectation \cite{Pol-Sus-Tho-Ugl} that in a completely local field
theory one cannot have complementarity.

We would like to mention in this context that the AdS/CFT
correspondence appears to necessitate a closer look at Hawking's
original derivation of thermal radiation from a black hole
\cite{Haw75}. One of the crucial assumptions in \cite{Haw75} is that
the Hilbert spaces of observables on the future null infinity ${\cal
  I}_+$ and on the horizon ${\cal H}_+$ (see \fig{f-collapse}) are
independent of each other, implying thereby that the observables
belonging to these Hilbert spaces commute. Now, recall that these two
sets of observables correspond, respectively, to (a) objects outside
the event horizon that escape to infinity and (b) infalling matter.
The discussion in the previous paragraph suggests that the observables
outside and inside the horizon are not quite independent and should
not be regarded as mutually commuting sets. Although that discussion
was in the context of near-horizon (anti-de Sitter) geometries and
does not include the asymptotically flat part, it already makes the
assumption in \cite{Haw75} of mutually commuting observables far from
obvious.

\sgap

\ni\underbar{Acknowledgement}: I would like to thank Justin David,
Avinash Dhar and Spenta Wadia for discussion and collaboration on many
topics mentioned here. I would like to thank the organizers of the
ICTP Spring School 1999 for providing an opportunity to present a
four-lecture series which constitutes an early version of this
review. The participants of the ICTP School took part in several
lively and useful discussions; I express my thanks also to them.  I
would also like to thank CERN theory division for hospitality during
the preparation of the present version of the review.

\end{document}